\documentclass{statsoc}
\usepackage[a4paper]{geometry}
\usepackage[textwidth=8em,textsize=small]{todonotes}
\usepackage{color,soul}
\usepackage[utf8]{inputenc}
\usepackage{ctable}
\usepackage[colorlinks=TRUE,linkcolor=blue,citecolor=blue]{hyperref}
\usepackage{float}
\usepackage{amsmath,amsfonts, bm} 
\usepackage{graphicx,psfrag,epsf}
\usepackage{enumerate}
\usepackage{mathtools}
 \usepackage{algpseudocode}
 \usepackage{algorithm}
\usepackage{natbib}
\usepackage{pdfpages}
\usepackage{placeins}
\usepackage{lscape}
\usepackage{textcomp}
\usepackage{xr-hyper}
\usepackage{hyperref}

\usepackage[normalem]{ulem}
\newtheorem{assumption}{Assumption}
\title[Direct and spillover effects. An SCM apporach]{Direct and spillover effects of a new tramway line on the
	commercial vitality of peripheral streets.
	A synthetic-control approach }
	
\author{Giulio Grossi}
\address{Department of Statistics, Computer Science, Applications, University of Florence,
Florence,
Italy.}
\email{giulio.grossi@unifi.it}

\author{Marco Mariani}
\address{IRPET - Istituto Regionale Programmazione Economica Toscana,
Florence,
Italy.}
\author{Alessandra Mattei}
\address{Department of Statistics, Computer Science, Applications, University of Florence,
Florence,
Italy.}
\author{Patrizia Lattarulo}
\address{IRPET - Istituto Regionale Programmazione Economica Toscana,
Florence,
Italy.}
\author[Grossi {\it et al.}]{\"{O}zge \"{O}ner}
\address{Department of Land Economy, Cambridge University,
Cambdridge,
UK.}

\numberwithin{equation}{section} 
\numberwithin{figure}{subsection} 
\numberwithin{table}{subsection} 

\newcommand*{\N}{{\cal{N}}}

\newcommand*{\bw}{{\mathbf{w}}}
\newcommand*{\be}{{\mathbf{e}}}
\newcommand*{\bz}{{\mathbf{0}}}

\newcommand*{\bC}{{\mathbf{C}}}

\newcommand*{\bD}{{\mathbf{D}}}

\newcommand\blfootnote[1]{%
  \begingroup
  \renewcommand\thefootnote{}\footnote{#1}%
  \addtocounter{footnote}{-1}%
  \endgroup
}

\begin{document}
\maketitle
\blfootnote{We wish to thank Fabrizia Mealli, Georgia Papadogeorgou, anonymous referees and editors for their extremely insightful comments and suggestions}
 
\begin{abstract}
\noindent In cities, the creation of public transport infrastructure such as light rails can cause changes on a very detailed spatial scale, with different stories unfolding next to each other within the same urban neighborhood. We study the direct effect of a light rail line built in Florence (Italy) on the retail density of the street where it was built and its spillover effect on other streets in the treated street's neighborhood. To this aim, we investigate the use of the Synthetic Control Group (SCG) methods in panel comparative case studies where interference between the treated and the untreated units is plausible, an issue still little researched in the SCG methodological literature.  We frame our discussion in the potential outcomes approach. Under a partial interference assumption, we formally define relevant direct and spillover causal effects. 

\end{abstract}



\newpage
\section{Introduction} 
Synthetic Control Group (SCG) methods  \citep{Abadie:2003, Abadie:2010, Abadie:2015} are an increasingly popular approach used
to draw causal inference under the potential outcome framework \cite[e.g.,][]{Rubin:1974}  
in panel comparative case studies. In these studies, the outcome of interest is observed for a limited number of treated units, often only a single one, and for a number of control units, with respect to a number of periods both before and after the assignment of the
treatment.
SCG methods focus on causal effects for treated units: 
for each point in time after the assignment of the treatment,  a weighted average of the observed potential outcomes of control units is used  to reconstruct the potential outcomes under control for treated units.   These  weighted averages are  named synthetic controls. The vector of weights is chosen by minimizing some distance between pre-treatment outcomes and covariates for the treated units and the  weighted average of pre-treatment outcomes and covariates for the control units.  See \cite{Abadie:2021} for a review of  the empirical and methodological aspects of SCG methods.

In the last two decades, SCG methods  have gained widespread popularity, and there has been a growing number of studies applying them to the investigation of the economic effects on particular locations of a wide range of events or interventions  
Initially, SCG methods  have been used in  panel studies where the outcome of interest is observed for a single treated unit \cite[e.g.,][]{Abadie:2003, Abadie:2010, Abadie:2015}.
Recently, they have been generalized to draw causal inference in panel studies where focus is on the average causal effects for multiple treated units \citep{Cavallo:2013, Acemoglu:2016, Gobillon:2016, Kreif:2016, Lhour:2021}.
Additional important theoretical and conceptual contributions include the comparison of SCG methods  with alternative approaches for program evaluation, the definition of synthetic control units, and the development of new estimators \citep{Doudchenko:2016, Xu:2017, Athey:2018, Bottmer:2021}. 

In this  methodological and applied causal inference literature, SCG methods have been implemented  using the potential outcome approach  under the Stable Unit Treatment Value Assumption (SUTVA), which rules out the presence of interference and hidden versions of treatments  \cite{Rubin:1980}.
The no-interference component of SUTVA, which states
that the treatment received by one unit does not affect the
outcomes of any other unit, may be arguable in many studies, where the events or interventions of interest may produce their effect not only on the units that are exposed to them (direct effects), but also on other unexposed units (spillover effects).
In the presence of interference, both scientists and policy makers may be interested not only in the direct effect of an intervention on the unit(s) where it actually takes place but also in the effects that the same intervention may have -- though in an indirect fashion -- on other units not exposed to the intervention. Therefore, disentangling  direct and spillover effects becomes the key objective of the analysis. However, the presence of interference entails a violation of the SUTVA and makes causal inference particularly challenging. 

Over the last years, causal inference in the presence of interference has been a fertile area of research.
Important theoretical works have dealt with the formal definition of direct and spillover effects and with the development of design and inferential strategies to conduct causal inference under various types of interference mechanisms, in both randomized and observational studies \cite[e.g.,][]{Hong:2006, Sobel:2006, Hudgens:2008,  Arpino:2016,   Forastiere:2018,   Papadogeorgou:2018, Huber:2021}. 
Despite such increasing interest, to the best of our knowledge, 
only the recent works by \cite{Cao:2019}  and \cite{Mellace:2020} deal with the application of synthetic control methods to comparative case studies where the no-interference assumption is not plausible.
In particular, \cite{Cao:2019} introduce -- 	under the assumption that  spillover effects are linear in some unknown parameter -- estimators for both direct treatment effects and spillover effects. They also investigate their asymptotic properties when the number of pre-treatment periods goes to infinity.  \cite{Mellace:2020}  introduce a procedure, called ``inclusive synthetic control method'', under which direct and spillover effects can be estimated using  control units potentially affected by spillovers. 

Motivated by the evaluation of the causal effects  of a new light rail line recently built in Florence (Italy) on the commercial vitality of the surrounding area, we propose to  contribute to the nascent literature on the use of the SCG approach in a setting with interference. To that end, our paper  makes both methodological and substantive contributions.

From a methodological perspective, we formally define direct and spillover effects in comparative studies where the outcome of interest is observed for a single  treated unit and a number of control units, for a number of periods before and after the assignment of the treatment.
The direct effect measures the effect of the intervention on the treated unit. Spillover effects measures
the effects of the treatment on untreated units belonging to the treated unit's neighborhood. 
These causal estimands are defined under  a
partial interference assumption \citep{Sobel:2006}, which states that interference takes place between units located near each other, but not between units that are sufficiently far away from one another.
Under partial interference, we use the penalized SCG estimator recently developed by \cite{Lhour:2021} to estimate  direct effects and spillover effects of the first type by exploiting information on control units that do not belong to the treated unit's neighborhood.

From a substantive perspective, we assess the direct effect of a new light rail line built in Florence (Italy) on the retail density of the street where it was built and its spillover on neighboring streets. 
We measure the retail density of a street using the number  of  stores every five hundred metres. 
This kind of application is original with respect to the previous field literature, which has often examined whether the creation of urban rail infrastructure is accompanied by changes in real estate values or gentrification of the area \cite[e.g.,][]{Cervero:1993, Baum:2000,  Bowes:2001, Kahn:2007, Pagliara:2011, Grube:2015, Budiakivska:2018, Delmelle:2019} 
and, only more seldom, whether it is accompanied by a higher firm density \citep{Mejia:2012, Pogonyi:2021} or by the settlement of new retailers  \citep{Schuetz:2015, Credit:2018}.
Moreover, it is worth noting that not all these empirical studies are fully embedded in an explicit causal framework, and that none of them addresses the issue of spillovers.

The paper is organized as follows. Section \ref{sec:application} describes the application that motivates the methodological development we propose and the available data. Section \ref{sec:methodology} presents the methodology. In Section \ref{sec:results}, we discuss how the methodology is applied to study the case of the Florentine light rail and present the results of the analysis. Section \ref{sec:conclusion} concludes the paper.

\section{Motivating application and related data}
\label{sec:application}
\subsection{A new light rail in Florence, Italy}
In addition to being a renowned art capital, Florence is also a city with nearly 400,000 residents and the hub of a wide commuting area. Away from the artworks and the pedestrian footpaths packed with store windows in the city center, the thoroughfares of peripheral Florence are often congested with cars.
From the early 1900s, the city of Florence developed an extensive public tram network on street-running tracks. Such a network was dismissed in 1958 in favor of public bus transport. 
In the following decades, the city of Florence suffered from soaring private motor vehicle transport, which led to congested traffic and undermined both the effectiveness and the attractiveness of public transport.  
In order to face these issues, the project of a new light rail network has been discussed for a long time, in a climate of doubt about the possibility of raising the necessary funds for the work. Moreover, there has been a strong debate about the appropriateness of this solution compared to others, also in view of the discomfort and discontent that long-lasting construction sites would have created in the areas exposed to the intervention. Nevertheless, a tram network project took shape during the 1990s.

The planned network mostly runs on reserved tracks, thus guaranteeing a more reliable public transport service, especially on long-distance journeys. Once completed, it will develop radially from the city center towards all the main surrounding suburbs. 

In the everyday slang of Florentines, the brand new light rail continues to be referred to by the old-fashioned term ``tramway.'' The first tramway line of the  network was constructed between 2006 and 2010. It connects the main railway station, in the city center, with the Southwestern urban area. 
The most intensive phase of works, when tracks were laid and stations were built, started in 2007. The first line
was completed in 2010. It has a total length of 7.6 kilometres, with stops approximately every 400 metres. After the inauguration of this line, some previous long-distance bus services were suppressed, whereas other ones were re-designed as short-distance services to ease access to the tramway from adjacent areas.
The completion of the planned light rail network requires the construction of four additional lines. The construction of two of these lines started in 2014 and was completed in 2018, while the remaining two lines are at a very preliminary stage. 
The analysis in this paper looks at the 1996-2014 period and focuses on the first line of the tramway. In particular, we consider the section of the line that goes along Talenti St.~(1.2 kilometres, 3 stops: Talenti, Batoni, and Sansovino), one of the main thoroughfares in the densely inhabited Soutwestern urban neighborhood of Legnaia-Isolotto (Legnaia hereinafter). There are other important thoroughfares and streets in Legnaia, most of which run parallel to Talenti St.~but do not host light rail tracks and stations. They are: Pollaiolo St.~(about 300 metres from Talenti St.); Pisana St.~(450 metres); Baccio da Montelupo St.~(500 metres), Scandicci St.~(650 metres); and Magnolie St.~(650 metres).  
For each of these streets, we consider a section of a maximum length of 1.2 kilometres, which we select to be  geographically the closest  to Talenti St. All these streets fall within 800 metres range  from the light rail and its transit stations (corresponding to a walking distance of about 10 minutes), which is considered a reasonable area of impact by the field literature \citep{Guerra:2012}. 
It is worth noting that, unlike previous studies, where streets within a given radius from transit infrastructures are aggregated to form a cluster-level unit, we consider each street as a distinct statistical unit.

\subsection{Conjectures on how light rail could affect the streets' retail activity}

Light rail is generally expected to raise accessibility through the improvement of transit times between different points within an urban area \cite[e.g., see][and the literature review therein]{Papa:2015}. However, citywide accessibility improvements are likely to occur in the presence of an extensive light rail network. This is not the case in our study, where there is only one light rail line, which was mainly conceived to make access to the city center easier from one particular section of the urban periphery. A single line like the one  in  our study is expected to yield a rather localized accessibility improvement.
At the same time, the light rail may be expected to trigger a process of revitalization of peripheral areas and of the retail sector therein. 
This may occur once the light rail is in operation thanks to high flows of transit users and renewed site image. However, the previous empirical literature suggests that the boost of the local retail sector, if any, can be small or transitory \citep{Mejia:2012, Schuetz:2015, Credit:2018}. 

Before the light rail inauguration, construction works may temporarily undermine the area's attractiveness and livability.
Faced with the light rail construction site in front of their shop windows, incumbent store owners often complain about the risk of lost opportunities owed to poor site image, traffic diversions, very limited street parking, and so forth. For the store owners located on other thoroughfares belonging to the same neighborhood of Talenti St., but with no construction site, the story might go the other way around during the tramway construction, with increased  opportunities owed to temporarily higher flows guaranteed by traffic diversions, unchanged image and street parking possibilities, increased relative competitiveness, and so forth.

When the new infrastructure goes into operation in a given site, the prospects of the commercial environment of adjacent sites are hard to envisage. On the one hand, they could also benefit from having the light rail at walking distance, which may increase the footfall for the retailers, constituting a positive spillover effect. On the other hand, they might return to business as usual, or even be crowded out and lose footfall due to the soaring relative attractiveness of the street where   stations are located, which may then constitute a negative spillover effect \citep{Credit:2018, Pogonyi:2021}.

The effect of the tramway on the commercial environment of a given shopping site may be heterogeneous depending on the different  types of stores. Since stores may belong to a high number of categories, an attractive way to group them into a few meaningful classes is to distinguish between purveyors of non-durable goods/frequent-use services (non-durables hereinafter) and purveyors of durable goods/seldom-use services (durables hereinafter). This distinction may help characterize in greater detail the effects of the light rail on an urban neighborhood's retail sector. Indeed, it reflects  a difference in the frequency of purchase of the two types of goods and services, which is very high for non-durables and relatively low for durables. It is also correlated with the customers' willingness to travel to purchase each type of goods and services: such willingness is generally low for non-durables, which are usually purchased in one's vicinity, and generally high for durables, which may see customers ready to bear some costs to patronize less accessible stores every once in a while \citep{Brown:1993, Klaesson:2014, Larsson:2014}.

\subsection{Data}
\begin{figure}
\caption{Stylized map of the neighbourhoods and the streets  involved in the analysis, clustered in their own urban neighborhoods}
\centering
\includegraphics[width=0.65\linewidth]{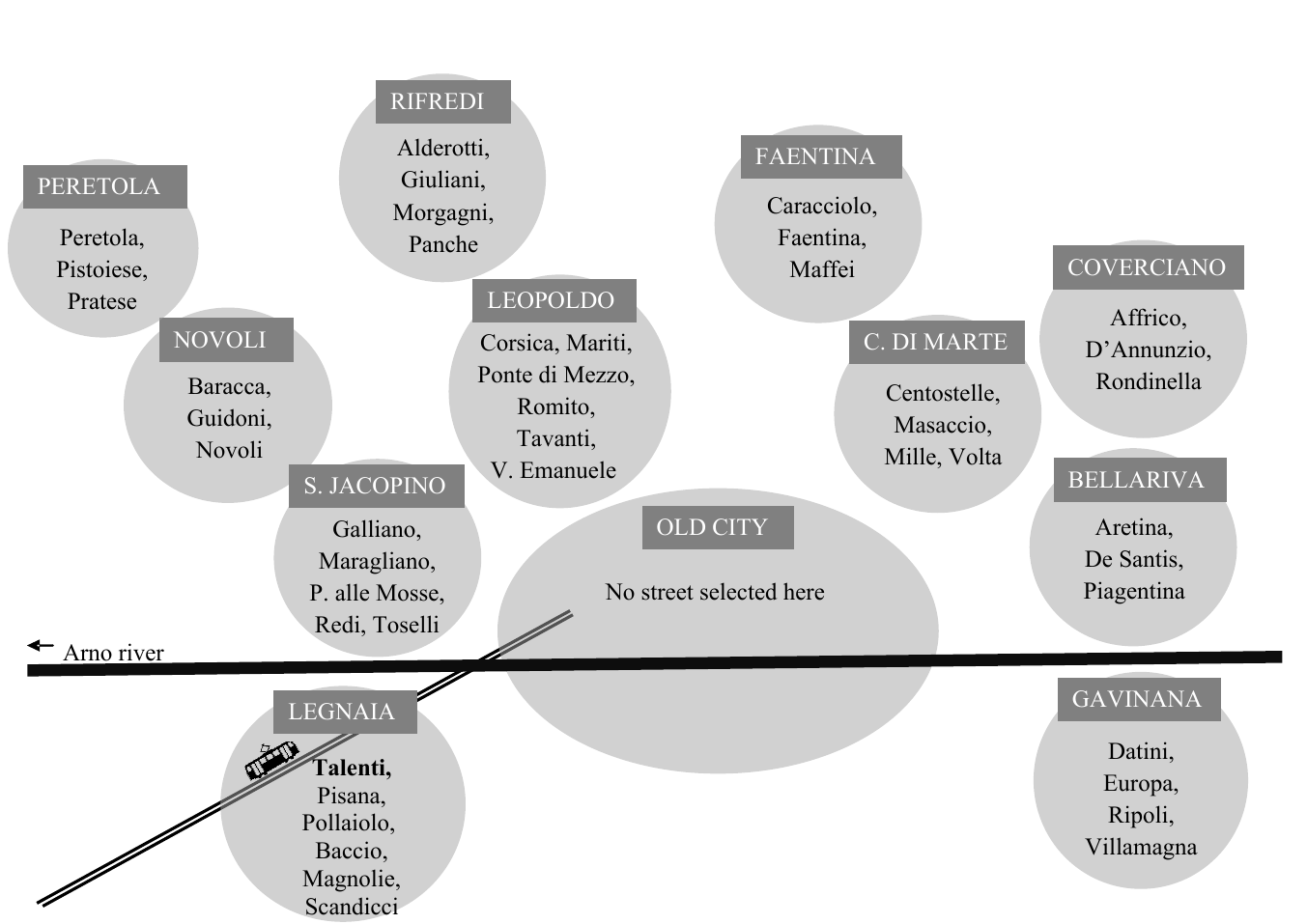}
\label{fig:map}
\end{figure}

The dataset used to examine the impact of the new light rail on the local retail environment includes  information on 6 streets in the peripheral urban neighborhood of Legnaia (Talenti St., Pisana St., Pollaiolo St., Baccio da Montelupo St., Scandicci St., and Magnolie St.) and on 38 further thoroughfares and streets of Florence, clustered in other 10 peripheral neighborhoods that are far from Legnaia. The definition of urban neighborhoods is based on the areas identified by the Real Estate Observatory of the Italian Ministry of Finance. We do not consider any street in the city center, as its commercial environment is completely different from what can be found in the surrounding residential neighborhoods.
A stylized map of the neighbourhoods and the streets, with their position in Florence, is shown in figure \ref{fig:map}. It is worth noting that the tramway passes only through the Legnaia neighborhoods and the city center, and the latter is excluded from the analysis. Moreover, Legnaia neighborhood is located in the south bank of Arno river, parted away from the other clusters by the river and the Bellosguardo hills. 
Background and outcome variables for each street originate from the Statistical Archive of Active Firms (SAAF, English translation of ASIA, the Italian acronym for ``Archivio Statistico delle Imprese Attive''). 
The SAAF is held by the Italian National Institute of Statistics (ISTAT). This dataset is available from 1996 onwards. It collects some basic, individual information on all the active local units of firms, including the exact location of the activity and the sector of activity (classified according to the Statistical Classification of Economic Activities in the European Community, usually referred to as NACE). We construct background and outcome variables for each street as follows. 
First, we select firms that are active in the retail sector in the city of Florence. Second, we further select only those stores having their shop windows on the streets involved in the study or that are located within an extremely short distance from such streets (50 metres). Third, in line with the reasoning developed in the previous subsection, we classify each of these stores into a NACE sector of activity in order to elicit the product/service these stores sell, and  group them into two categories: purveyors of durable goods (or seldom-use services); and purveyors of non-durable goods (or frequent-use services).
For each street and year, we finally construct background and outcome variables aggregating information across stores belonging to the same category. In our application, we focus on the following two outcome variables: number of purveyors of durable goods every 500 metres and; number of purveyors of non-durable goods every 500 metres, defined as $$Y_{it}=\frac{S_{it}}{L_i}\times 500$$ where $S_{it}$ is the number of durables/non-durables purveyors in street $i$ at time $t$, and $L_i$ is the length of street $i$.
\begin{figure}
\caption{Observed values of the  number of purveyors of durable (left panel)  and non-durable (right panel) goods every 500 meters over the time period 1996-2014 in the treated street (Talenti St.) and in other streets belonging to the same urban neighborhood (Pollaiolo St., Pisana St., Baccio da Montelupo St., Scandicci St., and Magnolie St.) }
	\centering
	\includegraphics[width=0.8\linewidth]{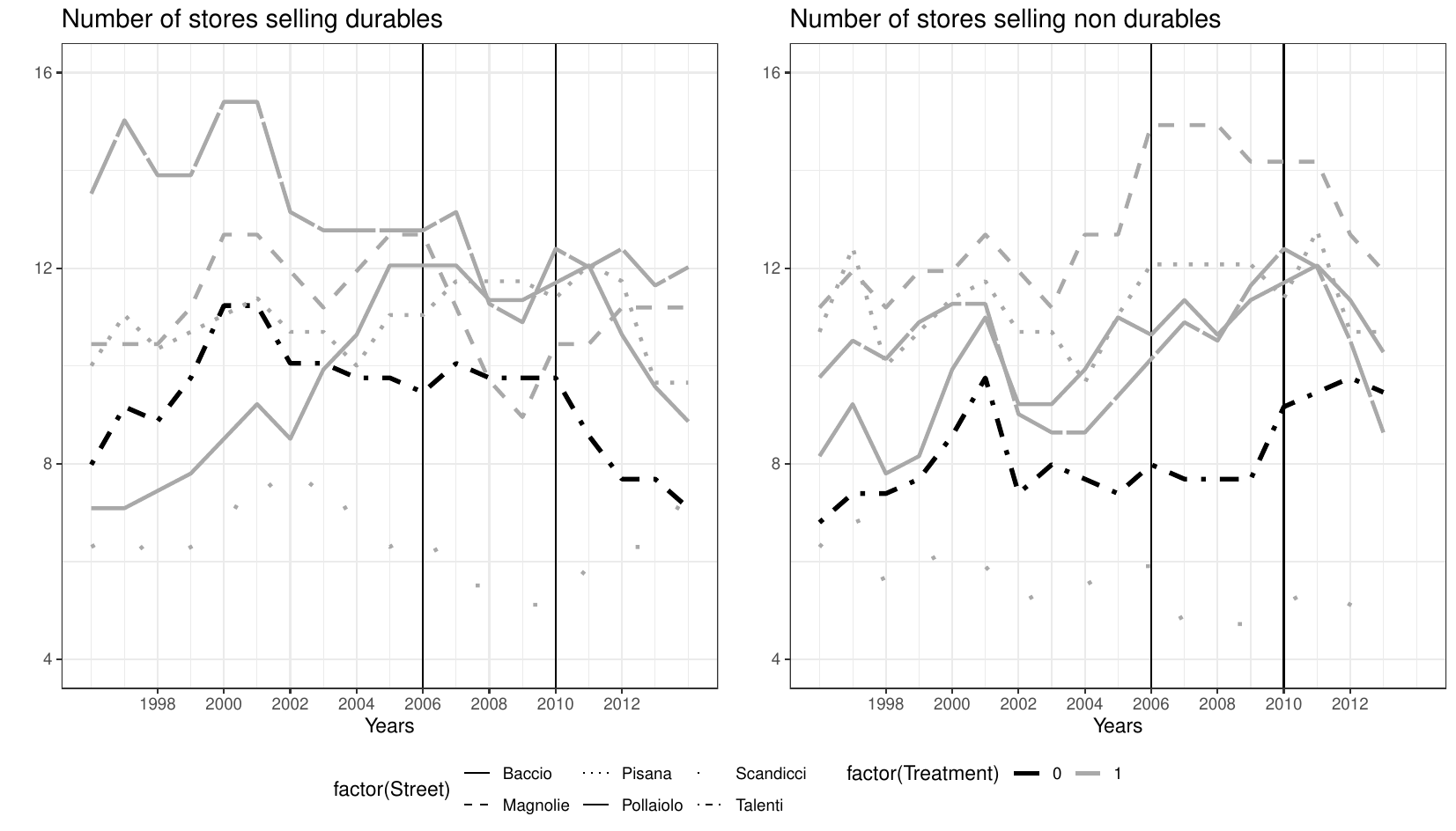}
\label{fig:legnaia4grafici}
\end{figure}
For each street, when focus is on number of stores selling durables, we use the following set of background variables consisting in 
\begin{itemize}
\item  The pre-treatment (1996–2005) time series of the number of stores selling durable goods every 500 metres
\item  The mean number of stores selling non-durables  averaged over the pre-treatment  1996–2004 period and 
the number of non-durable stores in 2005, the last pre-treatment year. 
\item A vector of time- and unit-specific covariates, representing information about the number of stores of the opposite kind (e.g.: information on stores selling non-durables, when we focus on durables selling stores)
\item  The mean number of stores selling durable in streets belonging to the same neighborhood averaged over the pre-treatment  1996–2004 period and the mean number of durable stores in streets belonging to the same neighborhood in 2005, the last pre-treatment year.
\end{itemize}
A similar set of background covariates is used when the focus is on the number of stores selling non-durables.  
Figure \ref{fig:legnaia4grafici} shows the observed value of the outcome  variables over the entire time period 1996-2014. The left-hand vertical line marks the start of light rail construction, and the right-hand vertical line marks the start of its operation.  As we are willing to examine the multi-faced impact of the tramway construction we will study its impact from 2006 to 2014, which is the period of active treatment.
The total number of stores (both selling durables and non durables goods), during the observation period, are shown in table \ref{tab:tot_stores}. We report the values for the street in the served neighborhood (Talenti, Pollaiolo, Pisana, Scandicci and Magnolie), their average excluding the served street and the average number of stores in the other streets of Florence. Similarly, table \ref{tab:tot_mean_stores} reports the number of stores every 500 metres, the main outcome. 

\begin{table}

\caption{Total number of stores in the Legnaia street(Talenti, Pollaiolo, Pisana, Scandicci and Magnolie), their average, excluded Talenti, and the average stores in the other streets in Florence}
\tiny
\centering
\resizebox{\textwidth}{!}{
\begin{tabular}[t]{lcccccccc}
\toprule
  & Talenti & Pollaiolo & Pisana & Scandicci & Magnolie & Baccio & Legnaia & Other streets\\
\midrule
1996 & 50 & 62 & 60 & 32 & 29 & 43 & 45.2 & 51.6\\
1997 & 56 & 68 & 68 & 35 & 30 & 46 & 49.4 & 53.9\\
1998 & 55 & 64 & 59 & 29 & 29 & 43 & 44.8 & 51.4\\
1999 & 59 & 66 & 62 & 31 & 31 & 45 & 47.0 & 56.3\\
2000 & 67 & 71 & 65 & 33 & 33 & 52 & 50.8 & 60.4\\
2001 & 71 & 71 & 67 & 35 & 34 & 57 & 52.8 & 63.3\\
2002 & 59 & 59 & 62 & 31 & 32 & 50 & 46.8 & 54.1\\
2003 & 61 & 57 & 62 & 34 & 30 & 54 & 47.4 & 53.2\\
2004 & 59 & 57 & 57 & 31 & 33 & 58 & 47.2 & 52.4\\
2005 & 58 & 59 & 64 & 31 & 34 & 65 & 50.6 & 55.6\\
2006 & 59 & 61 & 67 & 32 & 37 & 64 & 52.2 & 55.7\\
2007 & 60 & 64 & 69 & 26 & 35 & 66 & 52.0 & 55.3\\
2008 & 59 & 58 & 69 & 26 & 33 & 62 & 49.6 & 52.9\\
2009 & 59 & 60 & 69 & 25 & 31 & 64 & 49.8 & 52.2\\
2010 & 64 & 66 & 66 & 26 & 33 & 66 & 51.4 & 52.9\\
2011 & 61 & 64 & 72 & 29 & 33 & 68 & 53.2 & 53.7\\
2012 & 59 & 61 & 65 & 29 & 32 & 62 & 49.8 & 51.9\\
2013 & 58 & 54 & 59 & 28 & 31 & 56 & 45.6 & 47.3\\
2014 & 52 & 51 & 54 & 30 & 30 & 53 & 43.6 & 43.9\\
\bottomrule
\end{tabular}}
\label{tab:tot_stores}
\end{table}

\begin{table}

\caption{Stores every 500 metres in the Legnaia street(Talenti, Pollaiolo, Pisana, Scandicci and Magnolie), their average, excluded Talenti, and the average stores in the other streets in Florence}
\tiny
\centering
\resizebox{\textwidth}{!}{
\begin{tabular}[t]{lrrrrrrrr}
\toprule
  & Talenti & Pollaiolo & Pisana & Scandicci & Magnolie & Baccio & Legnaia & Other streets\\
\midrule
1996 & 14.8 & 23.3 & 20.7 & 12.6 & 21.6 & 15.2 & 18.7 & 21.6\\
1997 & 16.6 & 25.5 & 23.5 & 13.8 & 22.4 & 16.3 & 20.3 & 22.1\\
1998 & 16.3 & 24.0 & 20.4 & 11.4 & 21.6 & 15.2 & 18.5 & 21.0\\
1999 & 17.4 & 24.8 & 21.4 & 12.2 & 23.1 & 16.0 & 19.5 & 23.5\\
2000 & 19.8 & 26.7 & 22.4 & 13.0 & 24.6 & 18.4 & 21.0 & 24.8\\
2001 & 21.0 & 26.7 & 23.1 & 13.8 & 25.4 & 20.2 & 21.8 & 26.0\\
2002 & 17.4 & 22.2 & 21.4 & 12.2 & 23.9 & 17.7 & 19.5 & 22.9\\
2003 & 18.0 & 21.4 & 21.4 & 13.4 & 22.4 & 19.1 & 19.5 & 22.2\\
2004 & 17.4 & 21.4 & 19.7 & 12.2 & 24.6 & 20.6 & 19.7 & 21.9\\
2005 & 17.1 & 22.2 & 22.1 & 12.2 & 25.4 & 23.0 & 21.0 & 23.6\\
2006 & 17.4 & 22.9 & 23.1 & 12.6 & 27.6 & 22.7 & 21.8 & 23.2\\
2007 & 17.7 & 24.0 & 23.8 & 10.2 & 26.1 & 23.4 & 21.5 & 23.4\\
2008 & 17.4 & 21.8 & 23.8 & 10.2 & 24.6 & 22.0 & 20.5 & 22.1\\
2009 & 17.4 & 22.5 & 23.8 & 9.8 & 23.1 & 22.7 & 20.4 & 21.9\\
2010 & 18.9 & 24.8 & 22.8 & 10.2 & 24.6 & 23.4 & 21.2 & 22.2\\
2011 & 18.0 & 24.0 & 24.8 & 11.4 & 24.6 & 24.1 & 21.8 & 22.8\\
2012 & 17.4 & 22.9 & 22.4 & 11.4 & 23.9 & 22.0 & 20.5 & 21.6\\
2013 & 17.1 & 20.3 & 20.4 & 11.0 & 23.1 & 19.9 & 18.9 & 19.6\\
2014 & 15.4 & 19.2 & 18.6 & 11.8 & 22.4 & 18.8 & 18.2 & 18.4\\
\bottomrule
\end{tabular}}
\label{tab:tot_mean_stores}
\end{table}

These descriptive graphs suggest that, in Talenti St., the number of purveyors of non-durable goods (every 500 metres) increases after the tramway goes into operation.
On the other hand, the number of stores selling durables on Talenti St.~slightly increases during the early phase of construction but starts to diminish afterwards. 
On Pollaiolo St., the number of purveyors of non-durables grows during construction and wanes during the operational period. After an initial jump, Pisana St.~retains stores selling durables but loses some purveyors of non-durables when the light rail is operational.
Similarly, Baccio da Montelupo St. hosts a higher number of outlets during construction, followed by a later loss. 
On Scandicci St., the number of purveyors is overall stable. Finally, Magnolie St.~sees a continuous decrease in the number of stores selling durables, while the decline in the number of purveyors of non-durables begins as the light rail service starts.

\section{Methodology}\label{sec:methodology}
\subsection{Potential outcomes and observed outcomes} \label{sec:pos}
We consider a panel data setting with  $1+N$ units partitioned into $1+K$ clusters and observed in time periods $t = 1, \ldots, T$. Let $N_k$ be the number of units in cluster $k$, and let $\N_k$ denote the set of numbers indexing units that belong to cluster $k$, $k= 1, 2,\ldots, 1+K$.

In our motivating study, units are streets of Florence and clusters are naturally defined by  urban neighborhoods. Our dataset includes information  on  $1+N=1+43=44$ streets clustered into $1+K=1+10=11$  urban neighborhoods of Florence, which are  observed for $T=19$ years from 1996 to 2014.  
 Only Talenti St.~in the Legnaia neighborhood, say street $1$ in cluster $1$, 
is exposed to the intervention of interest: the construction of a new light rail line. 
The observed treatment period of nine years  comprises the four years (2006-2009) in which the tramway was built, and the last five years (2010-2014) when the tramway was operating, and thus we have $T_0=10$ pre-treatment years. 
In addition to Talenti St.,
the Legnaia neighborhood comprises  five  streets;  Pollaiolo St., Pisana St., Scandicci St., Magnolie St., and Baccio da Montelupo St.,  which we index by $i=2,3,4,5,6$, with $i \in \N_1$, respectively. 
The remaining 10 urban neighborhoods, which comprise 38 streets overall, are sufficiently far from Legnaia. See Figure \ref{fig:map} for a  stylized map.


Under the assumption that there is no hidden versions of treatment \cite[Consistency, ][]{Rubin:1980}, and the assumption of ``no-anticipation of the treatment''  \cite[e.g.][]{Abadie:2010, menchetti2022estimating}, let $$Y_{i,t}(\bw) \equiv Y_{i,t}([w_{1}, w_{2}, \ldots,  w_{1+N}]') \quad i \in \{1, \dots, N+1\}$$ denote 
the potential outcome for unit $i$ at time $t$ under  treatment assignment $\bw=[w_{1}, w_{2}, \ldots,  w_{1+N}]'$, where $w_i \in \{0,1\}$ for all $i=1, \ldots, 1+N$.

The assumption of ``no-anticipation of the treatment''  amounts to stating that the intervention has no effect on the outcome before the treatment period, $T_0+1, \ldots, T$, so that for $t=1, \ldots, T_0$, $Y_{i,t}([w_{1}, w_{2}, \ldots,  w_{1+N}]')=Y_{i,t}([0, 0, \ldots,  0]')$ for each unit $i=1, \ldots, 1+N$.



In this study, the  no anticipation of the treatment assumption appears to be plausible. In 2000,  the city administration announced the construction of the first line of the light rail network, but things soon turned out to be less easy than expected. The first tender for works attracted the interest of no construction companies. The outcome of the second call for tenders, in 2001, was the subject of a legal dispute lasting several years, giving rise to quite a few doubts -- in a public opinion that remained divided on the project -- as to whether and when a new light rail would ever exist in the city.  A third tender followed and the work was awarded to an unexpected consortium of those companies that had fought each other during the previous legal dispute. In light of such a troubled gestation, it is rather difficult to envision what kind of anticipatory behaviors, if any, might have been put in place by private economic agents, especially by the store owners that are the subject of the analysis proposed in the current paper.



When the population can be partitioned into clusters, it is often plausible to invoke the partial interference assumption  \citep{Sobel:2006}.
Such an assumption states that interference may occur
within, but not between, groups. 

\begin{assumption}
\label{ass:partial_interference}
(\textit{Partial Interference}).  Let $\bw_k=[w_i]'_{i \in \N_k}$ denote a treatment assignment sub-vector  for units in cluster $k$, $k=1, \ldots, 1+K$ and decompose 
$\bw \equiv  [\bw_1, \ldots, \bw_k, \ldots, $ $\bw_{K}, \bw_{1+K}]$. 
Then, for $t=T_0+1, \ldots, T$,  for all  
$$
\bw\equiv [\bw_1, \ldots, \bw_k, \ldots, \bw_{K}, \bw_{1+K}] \quad and \quad \bw^\ast\equiv 
[\bw^\ast_1, \ldots, \bw^\ast_k, \ldots, \bw^\ast_{K}, \bw^\ast_{1+K}]
$$
with $\bw_k=\bw^\ast_k$, $Y_{i,t}(\bw) = Y_{i,t}(\bw^\ast) $
for all $i \in \N_k$.
\end{assumption}

Partial interference implies that in each time period $t$ potential outcomes for unit $i$ in cluster $k$, $i \in \N_k$, only depend on its own treatment status and on the treatment statuses of the units belonging to the same cluster/neighborhood as unit $i$, but they do not depend on 
the treatment statuses of the units belonging to different clusters/neighborhoods. Therefore, partial interference allows us to write 
$Y_{i,t}(\bw)\equiv Y_{i,t}([\bw_{1}, \ldots, \bw_{k},\ldots, \bw_{1+K}])$ as 
$Y_{i,t}(\bw_{k})$ for all $i\in \N_k$ and   $t=T_0+1, \ldots, T$, for $k=1, \ldots, 1+K$.

In our application study, where streets are partitioned
into clusters defined by urban neighborhoods, it is rather plausible to assume that interference occurs within streets belonging to the same neighborhood, but not between streets belonging to different, geographically distant, urban neighborhoods. 
Indeed, we can reasonably expect that customers patronizing stores in a given peripheral area will hardly switch over to other distant, peripheral  areas because of a single light rail line connecting only one of these peripheries with the city center, but with none of the other peripheries.
Moreover, the treated cluster is naturally detached from the rest of Florence by the Arno River on the north side, and by the Bellosguardo hills on the East side, and we avoid considering control streets on its South and West side. Thus following, all the control neighbourhoods can be considered distant from the tramway line, and the spillover effects are unlikely to arise on them.

Throughout, let $Y_{i,t}$ denote the observed outcome. For the $T_0$ pre-treatment time periods, we observe all $1+N$ units without treatment, so that,
under consistency, no anticipation of treatment and partial interference, for $t=1, \ldots, T_0$,
$Y_{i,t}=Y_{i,t}(\bz_{N_k})$ for all $i\in \N_k$, $k=1,\ldots, 1+K$. 
In our study, Talenti St.~is the single treated unit, which we arbitrarily label as unit 1 in cluster 1, so for $t=T_0+1, \ldots, T$, the $T-T_0$ treatment time periods, we observe $Y_{i,t}= Y_{i,t}([1,\bz_{N_1-1}]')$,  for each $i\in \N_1$ and $Y_{i,t}= Y_{i,t}(\bz_{N_k})$, for each $i\in \N_k$, $k=2, \ldots 1+K$.

Throughout the paper, we refer to $Y_{i,t}(\bz_{N_1})$ for
$i \in  \N_1$ and $t=T_0+1, \ldots, T$ as control potential outcomes for the treated unit and for units who belong to the treated unit's cluster, respectively, and to  units that do not belong to the treated unit's cluster as control units.

\subsection{Causal estimands}\label{sec:estimands}
In a setting where only the first unit (Talenti St.) in the first cluster (Legnaia neighborhood) is exposed to the
intervention after time point $T_0$ (with $1 \leq T_0  < T$), and under the assumption of partial interference, we are interested in the following direct and spillover causal effects at time points $t=T_0+1, \ldots, T$.
  
We define the (individual) direct causal effect of treatment 1 versus treatment 0 for the treated unit/street as
\begin{equation}\label{eq:de}
\tau_{1,t}= Y_{1,t}(1,\bz_{N_1-1}) - Y_{1,t}(\bz_{N_1})\quad t=T_0+1, \ldots, T.
\end{equation}

 For all $i \in   \N_1\setminus \{1\}$, let 
$$\delta_{i,t} =Y_{i,t}(1,\bz_{N_1-1}) - Y_{i,t}(\bz_{N_1})$$ be the individual spillover causal effect of treatment 1 versus treatment 0 at time $t$ on unit $i$ belonging to cluster 1, the treated unit's cluster. 
We define the average spillover causal effect at time $t$ as
\begin{equation}\label{eq:ase}
\delta_{t}^{\N_1} = \dfrac{1}{N_1-1} \sum_{i \in \N_1 \setminus \{1\}} \delta_{i,t}  = \dfrac{1}{N_1-1}\sum_{i \in \N_1 \setminus \{1\}  }  \left[Y_{i,t}(1,\bz_{N_1-1}) - Y_{i,t}(\bz_{N_1})\right].
\end{equation}

Two remarks  on the causal effects we are interested in are in order. First, it is worth noting that we define direct and spillover effects as comparisons between potential outcomes under alternative cluster treatment vectors.
The literature on causal inference under partial interference has generally focused on average direct and spillover effects, defined as comparisons between average potential outcomes under alternative treatment allocation strategies \cite[e.g.,][]{Hudgens:2008, Papadogeorgou:2018}.
%
%
Second,  we are not interested in assessing causal effects for units/streets belonging to clusters/urban neighborhoods different from the treated unit's cluster (Legnaia), but the availability of information on them is essential for inference, as we will show in the next Sections.
We can re-write the 
(individual) direct causal effect for the treated unit in Equation \eqref{eq:de} and the average spillover causal effect in Equation \eqref{eq:ase} at time $t$, $t=T_0+1, \ldots, T$, as function of the observed outcomes:
$$
\tau_{1,t} =  Y_{1,t} - Y_{1,t}(\bz_{N_1}) \quad \hbox{ and } \quad 
\delta_{t}^{\N_1}  =  \dfrac{1}{N_1-1} \sum_{i \in \N_1 \setminus \{1\}}  \left[Y_{i,t} - Y_{i,t}(\bz_{N_1})\right].
$$
These relationships make it clear that we need to estimate   $Y_{i,t}(\bz_{N_1})$  for $i \in \N_1$ to get an estimate of $\tau_{1,t} $ and $\delta_{t}^{\N_1}$.  


\subsection{SCG estimators of direct and average spillover effects} \label{sec:SCG}
Under partial interference (Assumption~\ref{ass:partial_interference}), we creatively exploit information on units within clusters different from the treated unit's cluster to draw inferences on direct effects and average spillover effects using the SCG approach originally proposed by \cite{Abadie:2003, Abadie:2010}, and further developed by \cite{Lhour:2021}.

Several exiting SCG approaches exploit the idea of a stable relationship over time between the outcome of the treated units and the outcome of the control units in the absence of intervention \cite[stable patterns across units, e.g.,][]{Abadie:2003, Abadie:2010, Doudchenko:2016, Lhour:2021}. Similarly, our
method exploits stable patterns across units belonging to different clusters. Specifically, for each unit $i$ in cluster $1$, $i \in \N_1$, we assume that the relationship between the outcome of unit $i$, $Y_{i,t}$,	and the outcomes of  control units, $Y_{j,t}$, $j \in N_k$, $k=2, \ldots, 1+K$,
is stable over time. 
This type of stable pattern implies that: 
\begin{enumerate}
    \item  the same structural process drives both the outcomes of units in control clusters  (clusters of units that do not belong to the treated unit's cluster)  as well as the outcomes of the treated unit and its neighbors in the absence of treatment
    \item the outcomes of control units and their neighbors are not subject to structural shocks during the sample period of the study.
\end{enumerate}

Under  these assumptions, building on \cite{Abadie:2010}, we propose to impute the missing control potential outcomes for the treated unit and the units that belong to the treated unit's cluster as the weighted average of outcomes of control units. 

Formally, indexing the control units as $N_1+1,\ldots, 1+N$, for each unit $i$ in cluster 1, $i \in \N_{1}$, 
$$
\widehat{Y}_{i,t}(\bz_{N_1}) =\sum_{j= N_1+1}^{1+N}  \omega^{(i)}_{j} Y_{j,t}  \qquad t=T_0+1, \ldots, T,
$$ 
where $\omega^{(i)}_{j}$ are weights such that, for each 
$i \in \N_{1}$,
$$\omega^{(i)}_{j}\geq 0 \quad \hbox{for all }  j = N_1+1, \ldots, 1+N  $$ and $$\sum_{j = N_1+1}^{1+N}  \omega^{(i)}_{j}=1.$$
For each unit $i$ in cluster 1, $i \in \N_{1}$, the set of weights $\boldsymbol{\omega}^{(i)} = \left[ \omega^{(i)}_{N_1+1}, \ldots, 
 \omega^{(i)}_{1+N} 
 \right]'$ defines
the \textit{synthetic control unit} of unit $i$.

The choice of the weights, $\boldsymbol{\omega}^{(i)}$, is clearly an important step in SCMs.
The key idea is to construct synthetic controls that  best resemble the characteristics of the   units in the treated cluster before the intervention. Unfortunately,   the problem of finding a synthetic control that best
reproduces the characteristics of a unit may not have a unique solution. We face this challenge using the 
penalized synthetic control estimator recently developed by \cite{Lhour:2021}. 
In our setting, the penalized synthetic control estimator penalizes  pairwise discrepancies between
the characteristics of  units in the treated cluster and the characteristics of the units belonging to untreated clusters that
contribute to their synthetic controls. 


For each unit $i$, let $\bC_{i}=[C_{i,1}, \ldots, C_{i, H}]'$ be a $H-$dimensional vector of pre-treatment  individual covariates. Let
$Y^{(i)}_{\N_{k},t}= \sum_{j \in \N_k\setminus \{i\}} Y_{j,t}/(N_k-1)$, $t=1, \ldots, T_0$ and $C^{(i)}_{\N_{k}, h}=\sum_{j \in \N_k\setminus \{i\}} C_{j,h}/(N_k-1)$ denote the neighborhood-level pre-treatment outcomes and covariates for unit $i$ in cluster $k$, $k=1, \ldots, 1+K$.
Let $\bD_{i}=\left[Y_{i,1}, \dots, Y_{i,T_0}, Y^{(i)}_{\N_{k},1}, \dots, Y^{(i)}_{\N_{k},T_0}, \bC_{i}, \bC^{(i)}_{\N_{k}}\right]'$ denote the observed data 
for  unit $i$ in cluster $k$, $i \in \N_k$.

For each unit $i$ in the treated cluster $1$, 
the \textit{penalized} synthetic control vector of weights 
is chosen by solving the following optimization problem:
\begin{eqnarray}\label{eq:hatweights}
 \arg \min_{\boldsymbol{\omega}^{(i)}}
\left \| \bD_{i}  - \sum_{j = N_1+1}^{1+N}\bD_{j}\omega_{j}^{(i)}\right \|^2 + \lambda^{(i)}
\sum_{j = N_1+1}^{1+N} \omega_{j}^{(i)}\left \| \bD_{i}  - \bD_{j}  \right\|^2
\end{eqnarray}
  subject to $$\omega^{(i)}_{j}\geq 0 \quad \forall j \in N_1+1, \ldots, 1+N; \qquad \hbox{and} \qquad \sum_{j = N_1+1}^{1+N}\omega_{j}^{(i)}=1,$$
  where $\|\cdot \|$ is the $L^2-$norm: $\| \mathbf{v} \| = \sqrt{\mathbf{v}' \mathbf{v}}$ for   $\mathbf{v} \in \mathbb{R}^r$ and $\lambda^{(i)}$ is  a penalization constant.  \cite[see][for details on the costruction of the weights]{Lhour:2021}.   It is worth noting that the use of the $L^2$-norm implies that the same importance  is given to all  pre-treatment  individual- and neighborhood-level outcomes and covariates   as   predictors of the missing outcome.

  Under some regularity conditions,  if $\lambda^{(i)}$ is positive, then the optimization problem in Equation \eqref{eq:hatweights} has a unique solution \citep[see Theorem 1 in][]{Lhour:2021}. 
The penalization term 
defines a trade-off between   aggregate fit and   component-wise fit: the  penalized synthetic control estimator becomes   the synthetic control estimator originally introduced by \cite{Abadie:2003, Abadie:2010} as $\lambda^{(i)} \to 0$; and   the one-match nearest-neighbor matching with replacement estimator proposed by \cite{Abadie:2006}  as $\lambda^{(i)} \to \infty$. 


Given an estimate of the weights, $\widehat{\boldsymbol{\omega}}^{(i)}$, for each unit $i$ in the treated cluster $1$, we estimate the direct effects for the treated unit, $\tau_{1,t}$, and the average 
 spillover causal effects $\delta_{t}^{\N_1}$
 for $t=T_0+1,\ldots,T$, as follows:
\begin{equation}\label{eq:tauhat}
	\widehat{\tau}_{1,t} = Y_{1,t}  - \sum_{j = N_1+1}^{1+N}\widehat{\omega}_{j}^{(1)} Y_{j,t} 
\end{equation}
and
\begin{equation}\label{eq:deltahat}
	\widehat{\delta}_{t}^{\N_1} = \dfrac{1}{N_1-1} \sum_{i \in \N_1 \setminus \{1\}} \widehat{\delta}_{i,t} 
= \dfrac{1}{N_1-1} \sum_{i \in \N_1 \setminus \{1\}} \left[Y_{i,t} - 
\sum_{j = N_1+1}^{1+N}\widehat{\omega}^{(i)}_j Y_{j,t} \right].
\end{equation}

In the literature, various approaches  have been proposed to quantify uncertainty of SCG estimators,  both in the presence of a single treated unit as well as in the presence of multiple treated units. 
One of the most commonly used approaches uses falsification tests, also named ``placebo studies,'' \cite[][]{Abadie:2010, Abadie:2015, AndoSavje:2013,  Cavallo:2013, Acemoglu:2016, Firpo:2018}, but alternative approaches have been recently developed, which include the construction of conditional prediction intervals \cite[][]{Cattaneo:2021}, and conformal inference \cite[][]{Ben:2021, chernozhukov2021exact}. Finally, other works rely on re-sampling schemes such as wild bootstrap \citep{ben2022synthetic}, error resampling (\cite{Xu:2017}), or parametric bootstrap \citep{Arkhangelsky:2021}. 

None of these methods at the moment have been established as the standard for inference with synthetic control methods, and the debate on how quantifying uncertainty of SCG estimators is still vigorous.
In this paper, we opt for using an error resampling approach.
Specifically, we construct confidence intervals for the direct and average spillover effects, $\tau_{1, t}$ and $\delta_t^{\N_1}$, using the following residual resampling procedure. 

For each  control unit $i$, $i=N_1+1, \ldots, 1+N$,
we calculate the vector of errors as $$\widehat{e}_{i,t}= Y_{i,t} - \widehat{Y}_{i,t} \quad \text{ for }  t =1, \ldots, T,$$
where $\widehat{Y}_{i,t}$ is the synthetic-control estimate of $Y_{i,t}(0)$, which was obtained applying the method described above by using as control units all control units belonging to different clusters than unit $i$.
Let $\widehat{\pmb{e}}_i=\{\widehat{e}_{i,1}, \dots, \widehat{e}_{i,T}\}$ denote the vector of errors for  control unit $i$, $i=N_1+1, \ldots, 1+N$.

Under the assumption of error exchangeability for control units, we sample with replacement $1+N-N_1$ (the number of control units) error vectors from the set of error vectors
$\{\widehat{\pmb{e}}_{N_1+1} \ldots, \widehat{\pmb{e}}_{1+N}\}$ and  
  we can construct 'noisy'-pseudo outcomes for each control unit $i$, $i=N_1+1, \ldots, 1+N$, as
$$Y_{i,t}^{\ast}= \widehat{Y}_{i,t} + \widehat{e}^\ast_{i,t}, \quad t=1, \ldots, T.$$

For each unit $i$ in the treated cluster, $i \in \N_1=\{1,2, \ldots, N_1\}$, we  use the pseudo-pre-treatment outcomes, $Y_{i,t}^\ast$, $i=N_1+1, \ldots, 1+N$ and $t=1, \ldots, T$ to re-estimate its synthetic control outcomes, $\widehat{Y}_{i,t}^\ast$, $i \in \N_1$, $t=T_0+1, \ldots, T$ and thus, the direct and average spillover effects:
$\widehat{\tau}_{1,t}^{\ast}= Y_{1,t}- \widehat{Y}_{1,t}^\ast$, and 
$\widehat{\delta}_{t}^{\N_1,\ast}= \sum_{i \in \N_1 \setminus \{1\}} \left[Y_{i,t}- \widehat{Y}_{i,t}^\ast\right]/(N_1-1)$, for  $t =T_0+1, \ldots, T$.  

We repeat this procedure $B$ times to derive an estimate of the empirical distribution of the estimators of the direct effects, $\widehat{\tau}_{1,t}$, and  of the average spillover effects, $\widehat{\delta}_t^{\N_1}$, $t=T_0+1, \ldots, T$.
For  $t =T_0+1, \ldots, T$, let
$\widehat{\tau}_{1,t}^{\ast}(q)$ and $\widehat{\delta}_{t}^{\N_1,\ast}(q)$ denote the $q$ sample quantile of $\{\widehat{\tau}_{1,t}^{\ast, (1)}, \ldots, \widehat{\tau}_{1,t}^{\ast, (B)}\}$ and $\{\widehat{\delta}_{t}^{\N_1,\ast, (1)}, \ldots, \widehat{\delta}_{t}^{\N_1,\ast, (B)}\}$, respectively, 
Then, we construct pointwise bias-corrected  pivotal confidence intervals  for $\tau_{1,t}$  and $\delta_t^{\N_1}$, $t=T_0+1, \ldots, T$, as follows:
\begin{eqnarray}
\lefteqn{CI_{1-\alpha} \left(\tau_{1,t}\right)  =} \nonumber\\
&& \left( 2\widehat{\tau}_{1,t} - \widehat{\tau}_{1,t}^{\ast}(1-\alpha/2)-
 \mathrm{B}(\widehat{\tau}_{1,t}); 2\widehat{\tau}_{1,t} - \widehat{\tau}_{1,t}^{\ast}(\alpha/2)- \mathrm{B}(\widehat{\tau}_{1,t}) \right)
  \label{eq:ci_tau}
\\
\lefteqn{ CI_{1-\alpha} \left(\delta_{t}^{\N_1}\right)  =}\nonumber\\&&
\left( 2\widehat{\delta}_{t}^{\N_1} - \widehat{\delta}_{t}^{\N_1,\ast}(1-\alpha/2)-
 \mathrm{B}(\widehat{\delta}_{t}^{\N_1}); 2 \widehat{\delta}_{t}^{\N_1} - \widehat{\delta}_{t}^{\N_1,\ast}(\alpha/2)- \mathrm{B}(\widehat{\delta}_{t}^{\N_1}) \right)
\label{eq:ci_delta}
\end{eqnarray}
where the bias is calculated as difference between the estimate of the effects and the median of the estimated empirical distributions:
$\mathrm{B}(\widehat{\tau}_{1,t}) =\widehat{\tau}_{1,t} - \widehat{\tau}_{1,t}^{\ast}(q=0.5) $ and 
$\mathrm{B}(\widehat{\delta}_{t}^{\N_1})=\widehat{\delta}_{t}^{\N_1}-\widehat{\delta}_{t}^{\N_1,\ast}(q=0.5)$.




Alternative resampling schemes can be considered. In the web appendix, we consider three additional resampling schemes and show that they lead to 
the same inferential conclusions in our application study. Moreover, we perform classical placebo inference  finding comparable results (see the Web Appendix). 

\FloatBarrier
\section{Causal effects of a new light rail line on streets' retail density} 
\label{sec:results}

In this section, we apply the method described in Section \ref{sec:methodology} to estimate the direct, and the average spillover causal effects of a new light rail line on the retail sector density in a number of streets belonging to the same urban neighborhood in peripheral Florence (Italy). Talenti St., where the light rail is located, is subject to direct effects.
The nearby streets  -- namely Pollaiolo St., Pisana St., Baccio da Montelupo St., Scandicci St., and Magnolie St. -- may only be subject to spillovers originating from Talenti St.

The streets' retail density is measured using two street-level outcome variables: the number of stores selling durable and non-durable goods every 500 metres. We consider stores selling durable and non-durable goods separately because we  believe that effects can be heterogeneous for these two types of stores. Both the outcomes of interest were demeaned for the pre-treatment average outcome. 


\subsection{Penalized synthetic control estimators of direct and spillover effects}
\label{subsec1:results}

We  impute the missing potential outcomes  $Y_{i,t}(\bz_{N_1})$ for each $i \in  \N_1$ and  $t=T_0+1.\ldots, T$ applying the penalized synthetic control  method. 
For each street $i$ within the urban neighborhood of Legnaia, $i \in   \N_1$, we construct a synthetic street as  weighted average of other streets belonging to Florentine urban neighborhoods located sufficiently far away from Legnaia. 
From the imputed missing potential outcomes, we then estimate the direct and the average spillover causal effects of interest.

In order to estimate the penalized synthetic control weights following the procedure described in Section \ref{sec:SCG}, we primarily have to select an appropriate value  for $\lambda$.
 
In this work we use the leave-one-out cross-validation procedure proposed by \cite{Lhour:2021}.
First, for each post-intervention period $t=T_{0}+1,\dots, T$,  and  for each control unit $i$, $i=N_1+1,\ldots, 1+N$, we use the information on the  control  units expect unit $i$ to derive 
penalized synthetic control estimators of the potential outcomes  under control under different values of $\lambda$. 
Let $\widehat{Y}_{i,t}(\lambda)$ denote the penalized synthetic control estimator   of $Y_{i,t}(\bz_{N_k} )$  with penalty term $\lambda$, $i=N_1+1,\ldots, 1+N$, $t=T_0+1, \ldots, T$. 
For  each $t=T_{0}+1,\dots, T$, and $i=N_1+1,\ldots, 1+N$, we then calculate  
$$ Y_{i,t}- \widehat{Y}_{i,t}(\lambda) =  Y_{i,t}-\sum_{\underset{j\neq i}{j=N_1}}^{1+N}   w^{(i)}_{j}(\lambda) Y_{j,t}.$$
We choose    $\lambda$ to minimize the root mean squared prediction error (RMSPE) for the individual outcomes:
$$
\sqrt{\dfrac{1}{(T-T_0)} \sum_{i=N_1+1}^{1+N}  \sum_{t=T_0+1}^{T} \left[ Y_{i,t}-\widehat{Y}_{i,t}(\lambda)\right]^2}.
$$

In order to ensure the uniqueness and sparsity of the solution of the optimization problems in Equation \eqref{eq:hatweights}, we focus on values of  $\lambda \in (0, 1]$, testing  a total of 1000 values. 
Selected values for  $\lambda$ are reported in Table A2 of the web appendix 
Even if, in principle, the choice of $\lambda^{(i)}$ affects the final results, we can show that our results are robust to different choices of its value. Treatment estimation under different values of $\lambda^{(i)}$ is shown in the web appendix. 

Once we have selected the penalization term, we move to the calculation of the weights.
We estimate weights with the procedure described in \ref{sec:SCG}, using the covariates and the pre-treatment outcomes scaled with respect to the pre-treatment mean. 
The estimated weights are reported in Table A3 in web appendix. 

Given a value for $\lambda$ and the estimated  weights, $\widehat{\bm{\omega}}^{(i)}$, $i\in \N_1$,  we estimate direct effects, $\tau_{1,t}$, and average spillover effects, $\delta_{t}^{N_1}$, for $t=T_0+1, \ldots, T=2006, \ldots, 2014$, using Equations \eqref{eq:tauhat} and \eqref{eq:deltahat}.
The RMSPEs, calculated over the individual- and cluster-level pre-intervention outcomes for each street in Legnaia,  $i \in \N_1$, and its synthetic control, respectively, are reported in Table \ref{tab:rmspe}.
We derive 90\% pointwise biased corrected pivotal confidence intervals for these estimands using the residual resampling procedure described in section \ref{sec:estimands} with $B=1000$ replications. It is worth noting that in each replication, the estimates of the causal effects are derived using the penalized synthetic control method with the penalty term  $\lambda$   derived from the observed data.
We also assess the sensitivity of our inference to the choice of the parameter $\lambda$ by repeating the treatment effect estimations over the whole grid of $\lambda$. Results in the web appendix show the robustness of our estimates to the penalization parameter's choice. 
Results under alternative inferential resampling process, and under placebo inference approaches are reported as well in the web appendix. Results from these analyses do not change the inferential conclusion, suggesting any  sensitivity of our results to the modelling choices.



\subsection{Results}
\subsubsection{Estimated direct and average spillover effects}
\label{subsec2:de_ase}
\begin{figure}[t]
	\centering
\includegraphics[width=0.99\textwidth]{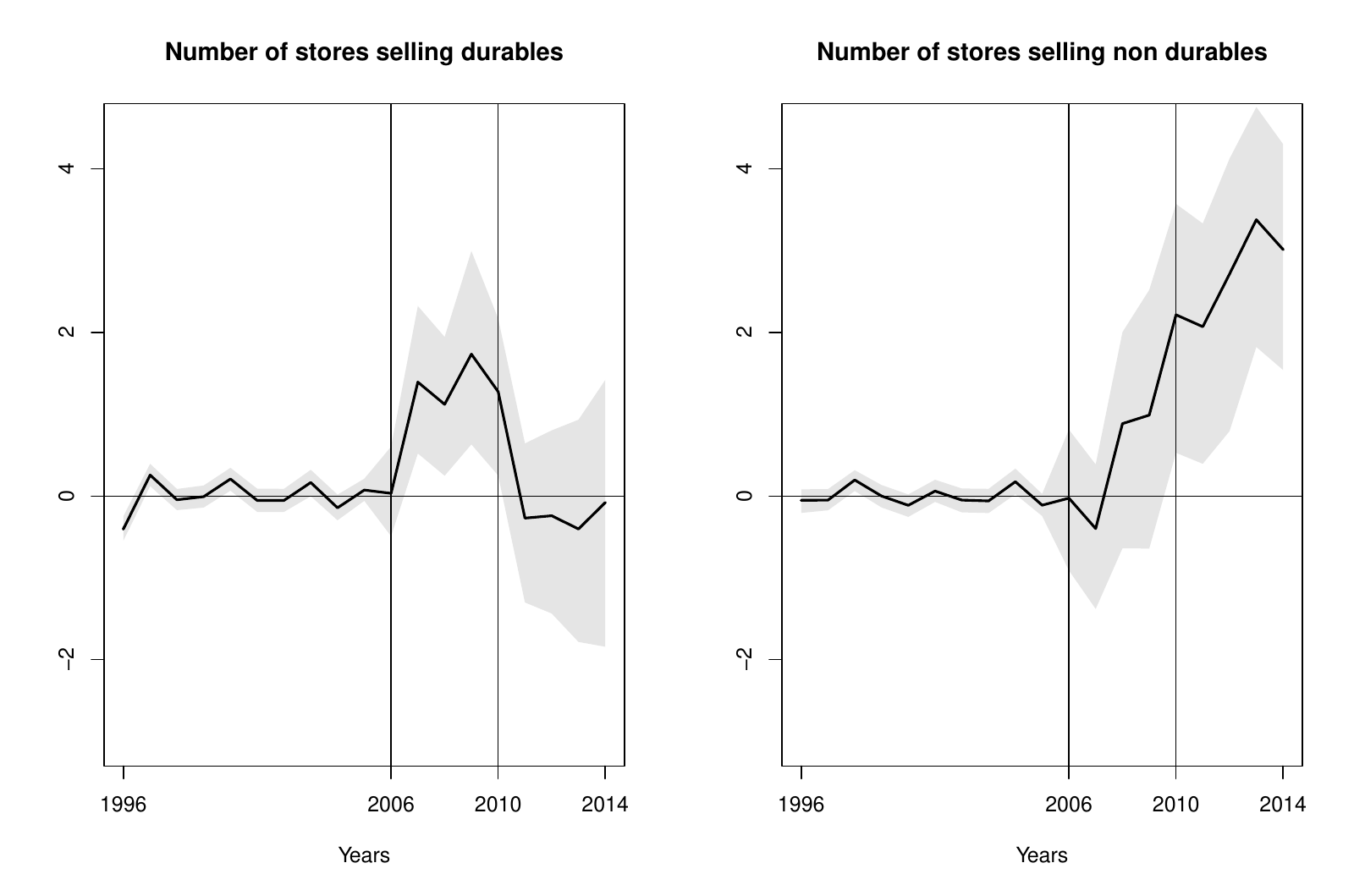}
	\caption{Estimated direct effects on Talenti St. (solid) and 90\% pointwise bias-corrected  pivotal confidence intervals (shaded area)}
	\label{fig:direct}
\end{figure}

Figure \ref{fig:direct} shows the estimated direct effect of the new light rail on Talenti St. 

\begin{table}
\caption{RMSPE  for the treated street, Talenti St. and the for untreated streets in the treated cluster}
\label{tab:rmspe}%
\centering
\begin{tabular}{|l|c|c|}
\hline
& \multicolumn{2}{|c|}{\textbf{$Y_{i,t}(1, \bz_{N_1-1})$}}\\
&\multicolumn{2}{|c|}{Number of stores selling} 	  \\
Street	&durable goods& non-durable goods\\
\hline
{Talenti St.}   & 0.1829& 0.1112\\
{Pollaiolo St.} & 0.2148& 0.1082\\
{Pisana St.}   &0.1854& 0.2305\\
{Scandicci St.}  &0.2452 & 0.2598\\
{Magnolie St.} &0.2080& 0.3024\\
{Baccio St.}  & 0.2619& 0.4416\\
\hline
\end{tabular}


\end{table}%
During the construction phase of the tramway (2006-2010), there is an increase in  the density of stores selling both durable and non-durable goods and the effects are statistically significant. During the operational phase of the tramway, however, the gain of durable goods purveyors  fades away, while the effect of the light rail remains positive, and of considerable magnitude, on the density of non-durable goods purveyors. 
A possible interpretation of these results is that the construction of the tramway initially beckons all types of retailers, who envision that the site will soon offer new commercial opportunities. However, increased demand should translate into higher prices for the available commercial space. Therefore, over a longer time horizon, purveyors of durables, which are goods with a lower frequency of purchase and higher customers' willingness to bear accessibility costs, have less incentive to pay the price required to stay next to the running tramway, because their customer base is not really made up of the occasional crowds of passers-by at stations. 
In contrast, purveyors of non-durable goods, which have a high frequency of purchase in one's vicinity, e.g. cafes, grocery stores, florists, and newsagents, depend more on these crowds of passers-by and, therefore, they are willing to pay the higher price required to stay on the site. These results are quite in line with the previous empirical literature, which highlights signs of commercial revitalization close to transit stations located in urban areas \citep{Credit:2018, Schuetz:2015}. 

\begin{figure}[t]
	\centering
\includegraphics[width=0.99\textwidth]{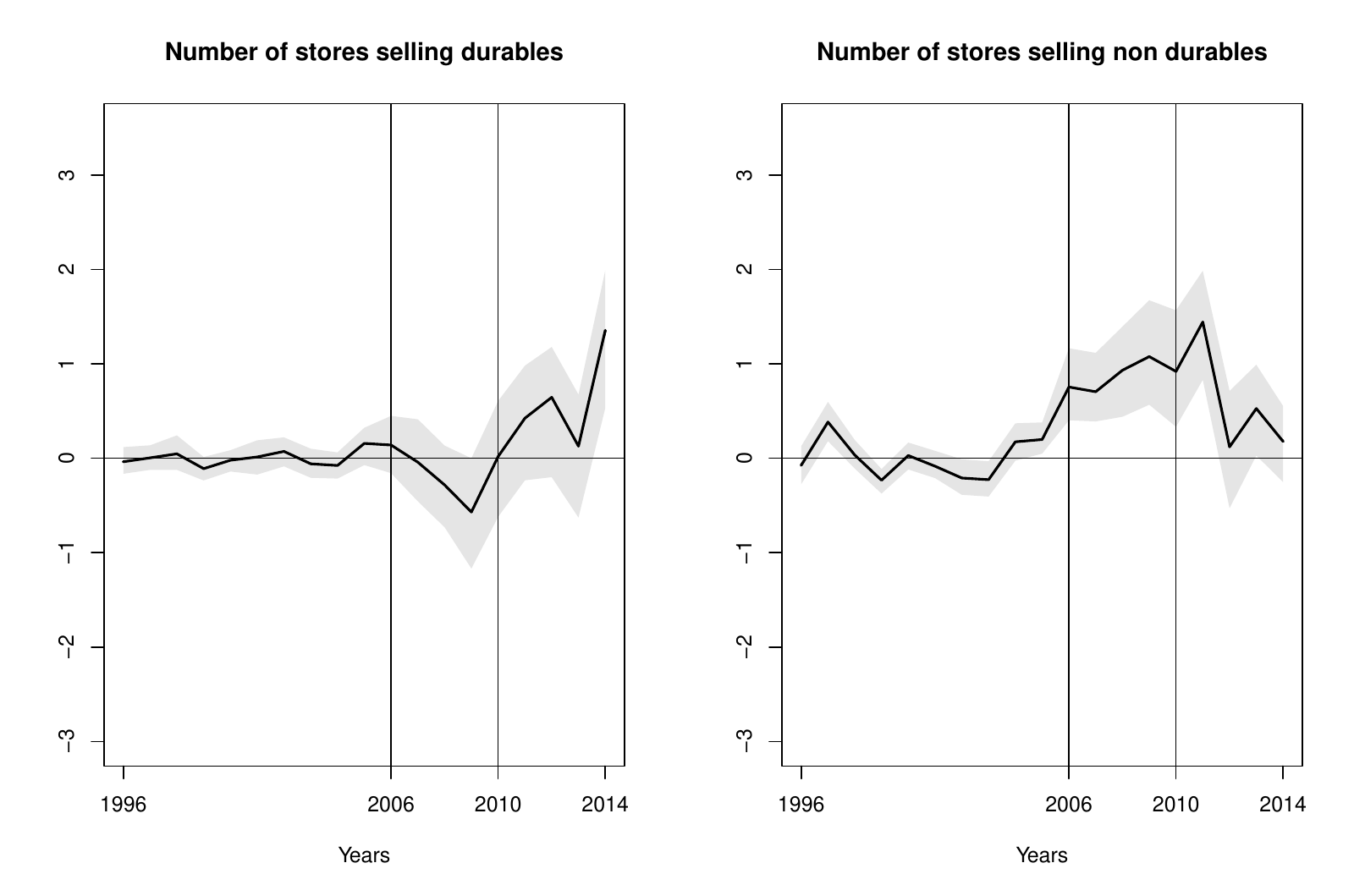}
\caption{Estimated average spillover effects on neighboring streets (solid) and 90\% pointwise bias-corrected  pivotal confidence intervals (shaded area)}
	\label{fig:spillover}
\end{figure}
The average spillover effects on the other streets in the urban neighborhood of Legnaia are shown in Figure \ref{fig:spillover}. 
As long as Talenti St. is undergoing construction works, we estimate   slightly negative effects on the density of durable goods retailers in the neighboring streets. Although these effects are not statistically significant, they confirm the idea that the construction of the tramway might have initially raised expectations about Talenti St. to the detriment of other commercial locations nearby. Then, after the light rail goes into service in 2010, the effect on the density of durable-goods purveyors in these alternative locations turns positive but small, as it is less than one store every 500 metres, and statistically negligible for most of the years.
Probably, for purveyors that depend little on occasional passers-by, shop windows on these streets are more worth their price than the coveted shop windows on Talenti St. Instead, with respect to stores selling non-durables, we have  positive and statistically significant effects on neighboring streets while the light rail is under construction in Talenti St., but such effects tends to fade and lose statistical significance afterward. A likely interpretation of this result is that, during construction, these alternative streets are expected to offer the opportunity to ``steal'' some of the customers that used to patronize stores selling non-durables on Talenti St., assuming that these customers would have been willing to flee the construction site to do their daily shopping within walking reach, or obliged to do so due to traffic detours. It is only a short-lived advantage, as Talenti St. later becomes the most lucrative place for non-durable goods purveyors due to the crowds coming and going all day at light rail stations.
In summary, the most noticeable quantitative effects occur in the street where light rail stations are located, as also found by the previous literature, but in the  streets close by there is no overt displacement. Rather, our results suggest that the tramway triggered divergent processes of commercial specialization: it strongly encourage the use of commercial spaces near stations by purveyors of non-durables, while it slightly increase the focus of other streets on the retail of durable goods. Highlighting these divergent specialization processes represents, in our view, an original contribution we make to the subject literature.

\section{Concluding Remarks}\label{sec:conclusion}

The SCG method has been hailed as ``\ldots the most important innovation in the policy evaluation literature in the last 15 years'' \citep{AtheyImbens:2017}  and the ideas initially put forward in
\cite{Abadie:2003, Abadie:2010, Abadie:2015}  have sparked avenues of methodological research.  This paper has met the challenge of extending the SCG method to settings where the assumption of interference is untenable. This is a nascent stream of research in the SCG literature, which our study contributes to inaugurate, with relevant implications for applied economic and social research.  

In this paper, building on recent methodological works on causal inference with interference in the potential outcomes framework, we  have formally defined unit-level direct effects and  average spillover causal effects under a partial interference assumption. We believe that these quantities may be relevant for a comprehensive evaluation of interventions at the meso- and macro-economic level.  
Then,  we have proposed to use the penalized SCG estimator \citep{Lhour:2021} 
to estimate direct and average spillover causal effects, capitalizing on the presence of clusters of units where no unit is exposed to the treatment. 

Our study has been motivated by the evaluation of the direct  effects of a new light rail line built in Florence, Italy, on the retail environment of the street where it was built, and of the spillover effects of the light rail on a number of streets close by. Although we focus on the Florence case study, similar interventions are often planned in other cities, too. Evaluating their  direct and spillover effects may provide precious insights to  policymakers,  helping them to understand what transformations in the urban landscape are being brought about by creating new transit infrastructure.
Our approach is very original also with respect to the field literature, where causal studies are still scarce and scholars usually conduct their analyses by aggregating all streets within a given radius (usually half a mile) from the new infrastructure.  From such a picture, we learn that the light rail has encouraged the emergence of divergent patterns of commercial specialization between the street hosting the stations with the crowds of passers-by, and the streets a little further away from the new light rail.

Our results rely on the assumption of partial interference, which is plausible in our application study,  as it is in many other causal studies \cite[e.g.,][]{Papadogeorgou:2018,  Huber:2021, Forastiere:2021}.  Nevertheless, we are aware that some studies might require a more general structure of interference \cite[e.g.,][]{Forastiere:2018, Forastiere:2021}. Therefore a valuable topic  for future research is the extensions of SCM methods to causal studies with general forms of interference.  

\renewcommand{\refname}{ \textbf{References}}

\bibliographystyle{apalike}

\newpage
\bibliographystyle{rss}
\bibliography{bibtex}

\section{Sensitivity Checks}

In this section, we wish to evaluate the robustness of some modeling choices that we have implemented. In particular, we assess the robustness of our results under different hypotheses for the interference across units and the robustness with respect to the choice of the penalization parameter $\lambda$. 

\subsection{Interference check}

In this subsection, we consider the alternative results for our analysis if we have tightened the partial interference assumption into the more classical no interference assumption across units. Under this hypothesis, the streets belonging to the treated neighbourhood would not receive any spillover effect and thus can be added in the donor pool.  

\begin{figure}
    \centering
    \includegraphics[width=.8\textwidth]{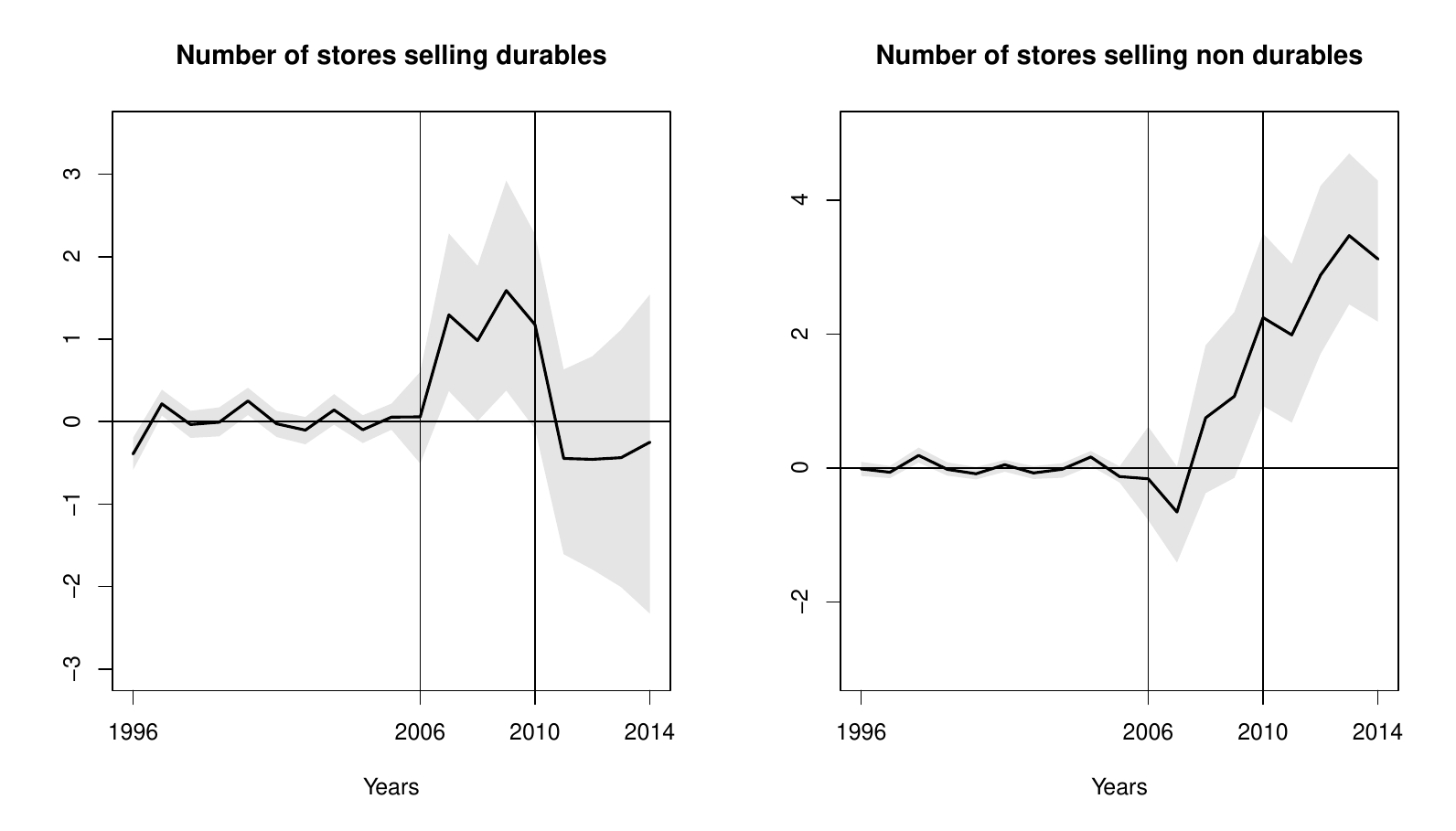}
    \caption{Direct effect estimation ignoring the interference across units - Shaded area: 95\% confidence intervals}
    \label{fig:interference_sens}
\end{figure}

As we can see, these results are very similar to the ones reported in the main text, under the assumption of partial interference. It seems that relaxing the no interference assumption has no particular effect on the estimation of the direct effect of the tramway.
\FloatBarrier
\subsection{Penalization term}

\begin{figure}[H]
        \centering
        \includegraphics[width=\textwidth]{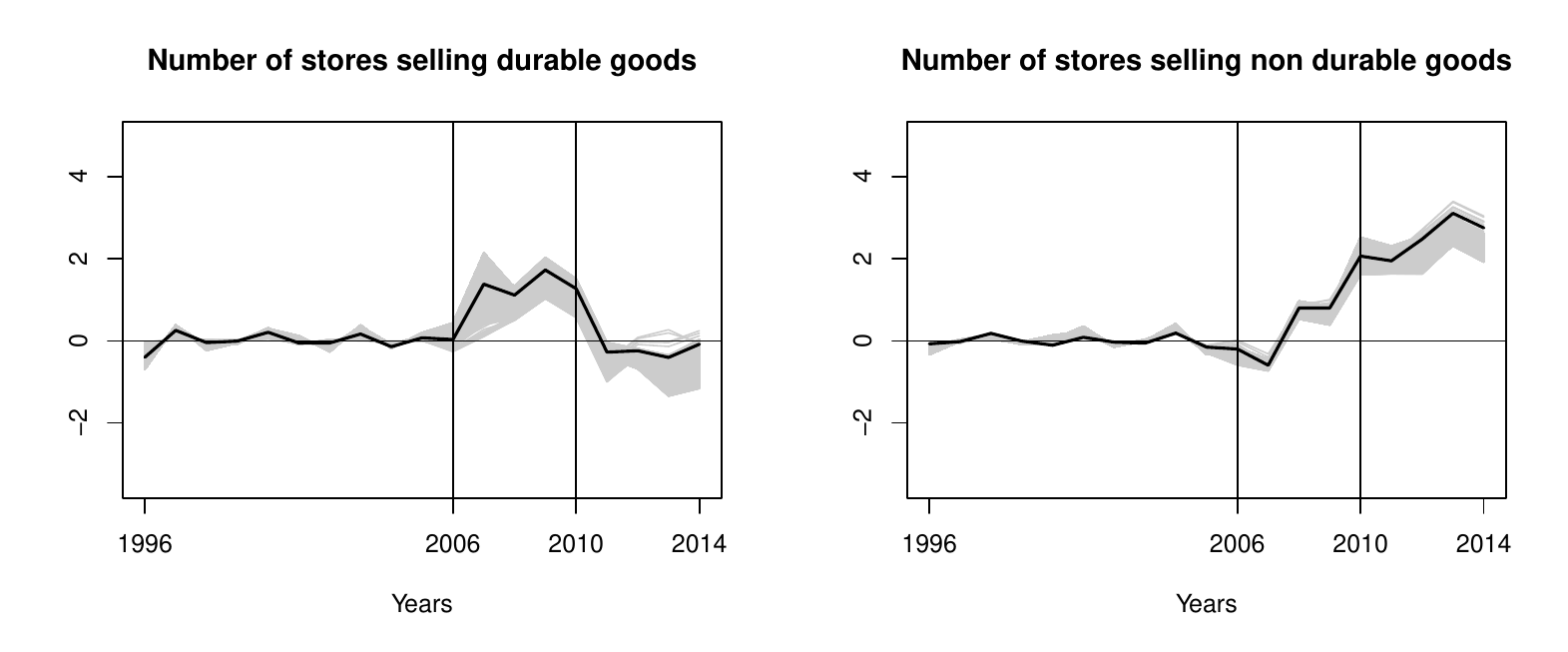}
        \caption{Direct Effect with different lambda values - Black line: Estimated direct effect - Grey lines: direct effect for alternative $\lambda$}
        \label{fig:l_ss}
    \end{figure}

\begin{figure}[H]
        \centering
        \includegraphics[width=\textwidth]{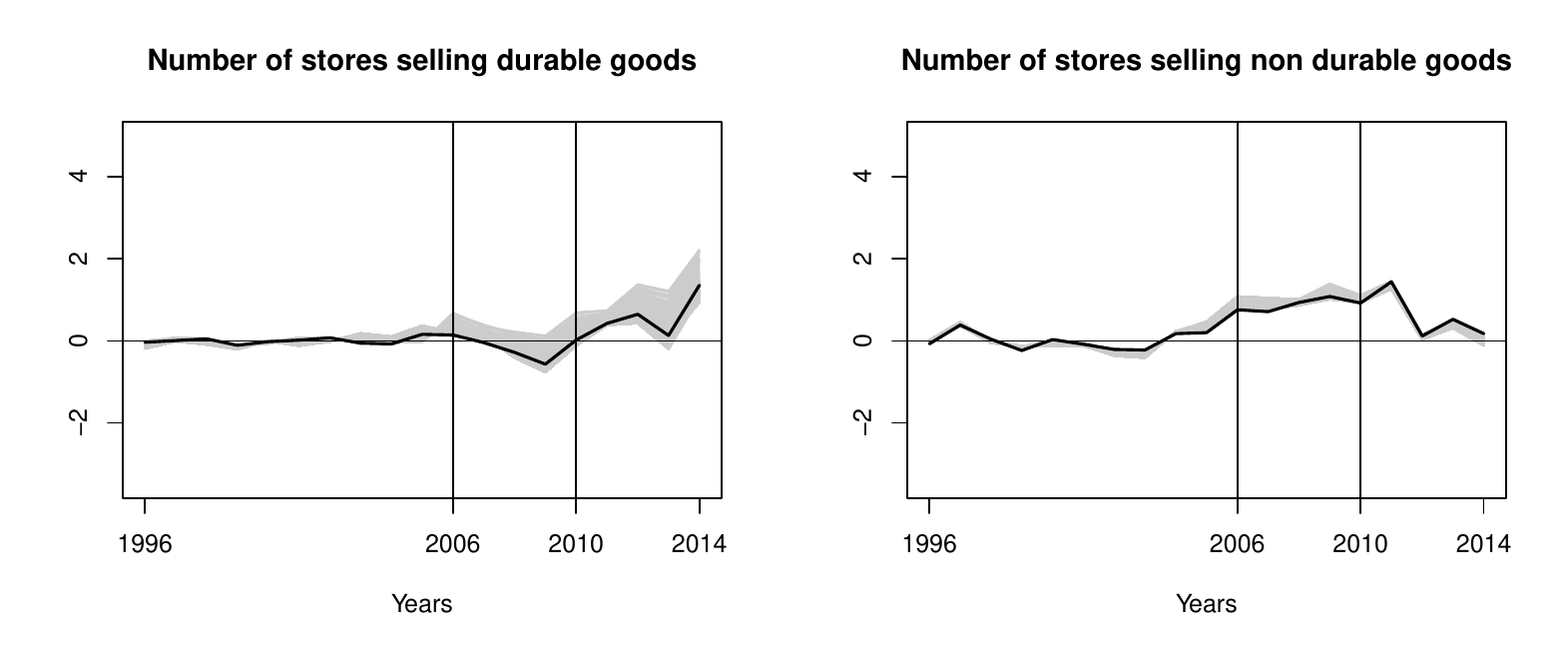}
        \caption{Average Spillover Effect with different lambda values - Black line: Estimated direct effect - Grey lines: direct effect for alternative $\lambda$}
        \label{fig:l_ss_avg}
    \end{figure}

In this subsection, we assess the sensitivity of our findings to different specifications of the penalization parameter $\lambda$. In principle, it is possible that the estimated effect varies dramatically when there is a minor shift in this parameter. Here in figures \ref{fig:l_ss} and \ref{fig:l_ss_avg}, we report the estimated direct effect for the whole grid of lambda involved in the analysis, from 0.001 to 1. As we can notice, the change in the penalization parameter does not correspond to a dramatic change in the value of the estimated effect. 

\FloatBarrier




\FloatBarrier

\section{Inference}

\subsection{Placebo tests}
To investigate the robustness of the results to the proposed inferential procedure we draw inferences on the causal effects of interest using various additional approaches.

First, we performed classical placebo tests  \cite{Abadie:2010}.
Specifically, we performed so-called in-space placebo studies. 
Following  the procedures described in \cite{Abadie:2010} and \cite{Abadie:2021}, we implement in-space placebo studies for 
the direct and spillover effects averaged over the construction period of the tramway (2006-2009) and over the period in which the tramway was operating (2010-2014):
$$
\tau_1^{c}=\dfrac{1}{4} \sum_{t=2006}^{2009} \tau_{1,t}\qquad 
\delta^{c, \N_1}=\dfrac{1}{4} \sum_{t=2006}^{2009} \delta_t^{\N_1}
$$
and
$$
\tau_1^{o}=\dfrac{1}{5} \sum_{t=2010}^{2014} \tau_{1,t}\qquad 
\delta^{o, \N_1}=\dfrac{1}{5} \sum_{t=2010}^{2014} \delta_t^{\N_1}
$$
where the superscripts $c$ and $o$ stand for ``construction" and operating period.
We use the ratio between the pre-and post-intervention root mean square prediction errors as test statistic.
The in-space placebo tests for $\tau_1^{o}$ and
$\tau_1^{c}$ are derived as follows. 
For $i=1, \ldots, N$, we calculate
$$\theta_i^c = \frac{RMSPE_i(2006:2009)}{RMSPE_i(1996:2005)}  \qquad  \hbox{and} \qquad
\theta_i^o = \frac{RMSPE_i(2010:2014)}{RMSPE_i(1996:2005)} 
$$

and derive the permutation distribution of $\theta_i^c$ and $\theta_i^o$ by iteratively reassigning the treatment to a unit in the control group and using all the other units except the untreated units in the treated cluster to construct the donor pool. Note that some control units were discarded from the analysis because of poor pre-treatment fit. 
The $p-$value for the inferential procedure based on the permutation distribution of $\theta_j^{d}$, $d=c,o$, is given by
 $$
 \frac{\sum_{j=N_1 +1}^N \mathbb{I}(\theta_j^d \geq \theta_1^d)}{N - N_1}\qquad d=c,o.
 $$
 The in-space placebo tests for $\delta_1^{o, \N_1}$ and $\delta_1^{c, N_1}$ are derived using a similar procedure on within-cluster average. 
For $k=1$, we calculate
$$\theta_{N_1}^c = \frac{RMSPE_{N_1}(2006:2009)}{RMSPE_{N_1}(1996:2005)}  \qquad  \hbox{and} \qquad
\theta_{N_1}^o = \frac{RMSPE_{N_1}(2010:2014)}{RMSPE_{N_1}(1996:2005)} 
$$
where the RMSPEs are calculated using as a response variable the average of the outcome values over the untreated units in the treated unit cluster. 
We then create $K^\ast=1000$ artificial clusters of size $N_1-1$ of control units sampling with replacement from the $N-N_1$ control units.
For $k=1, \ldots, K^\ast$, we calculate
$$\theta_{N_k}^c = \frac{RMSPE_{N_k}(2006:2009)}{RMSPE_{N_k}(1996:2005)}  \qquad  \hbox{and} \qquad
\theta_{N_k}^o = \frac{RMSPE_{N_k}(2010:2014)}{RMSPE_{N_k}(1996:2005)} 
$$
The $p-$value for the inferential procedure based on the permutation distribution of $\theta_{N_1}^d$, $d=c,o$ is given by
 $$
 \frac{\sum_{k=1}^{K^\ast} \mathbb{I}(\theta_{N_k}^d\geq \theta_{N_1}^d)}{K^\ast}\qquad d=c,o.
 $$
Table \ref{tab:placebo} shows the results.  
\begin{table}
      \caption{\label{tab:placebo}Estimates and p-values for direct and average spillover effects average over time}
      \centering
    \begin{tabular}{clcccc}
    \toprule
          &       & \multicolumn{2}{c}{Durable stores} & \multicolumn{2}{c}{Non durable stores} \\
          \midrule
          & Period & \multicolumn{1}{l}{Estimate} & \multicolumn{1}{l}{p-value} & \multicolumn{1}{l}{Estimate} & \multicolumn{1}{l}{p-value} \\
          \midrule
    $\tau_{1}$ & Construction & 1.382 & 0.025 & 0.336 & 0.225 \\
          & Operation & -0.234 & 0.705 & 2.514 & 0.023 \\
          \midrule
    $\delta_{N_1}$ & Construction & -0.188 & 0.139 & 0.883 & 0.135 \\
          & Operation & 0.516 & 0.031 & 0.704 & 0.234 \\
          \bottomrule
    \end{tabular}%
    
  \end{table}%
The results show positive and statistically significant direct effects on the number of stores selling durables during the construction period and on the number of stores selling non durables during the operational period of the tramway. We also find positive and statistically significant spillover effects 
on the number of stores selling durables during the operational period of the tramway. These results are consistent with those we find using the proposed error sampling approach.
    \begin{figure}
        \centering
        \includegraphics[width=.6\textwidth]{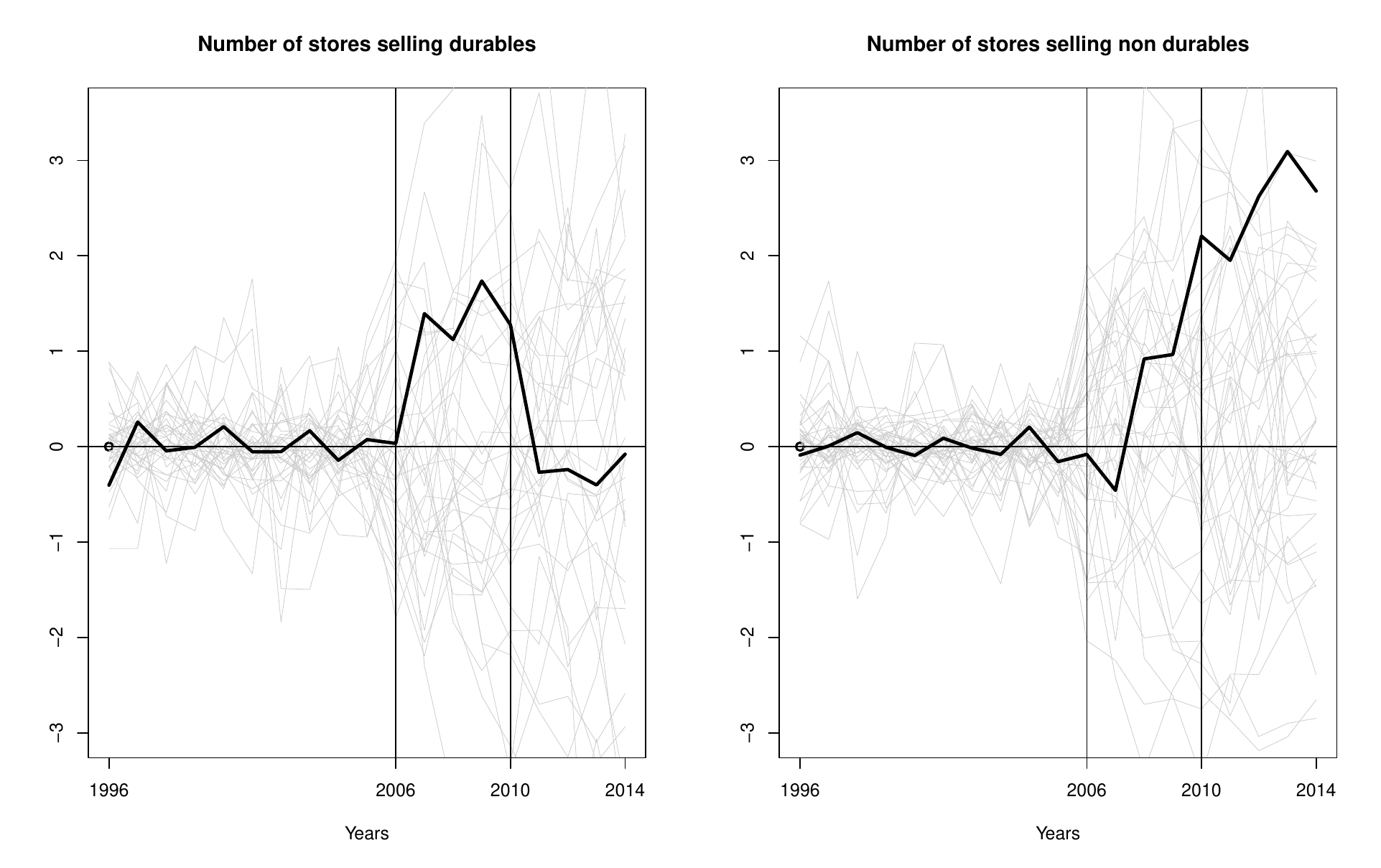}
        \caption{Placebo direct effects for stores selling durable and non durable goods - black line: Average spillover for Legnaia streets}
        \label{fig:enter-label}
    \end{figure}

        \begin{figure}
        \centering
        \includegraphics[width=.6\textwidth]{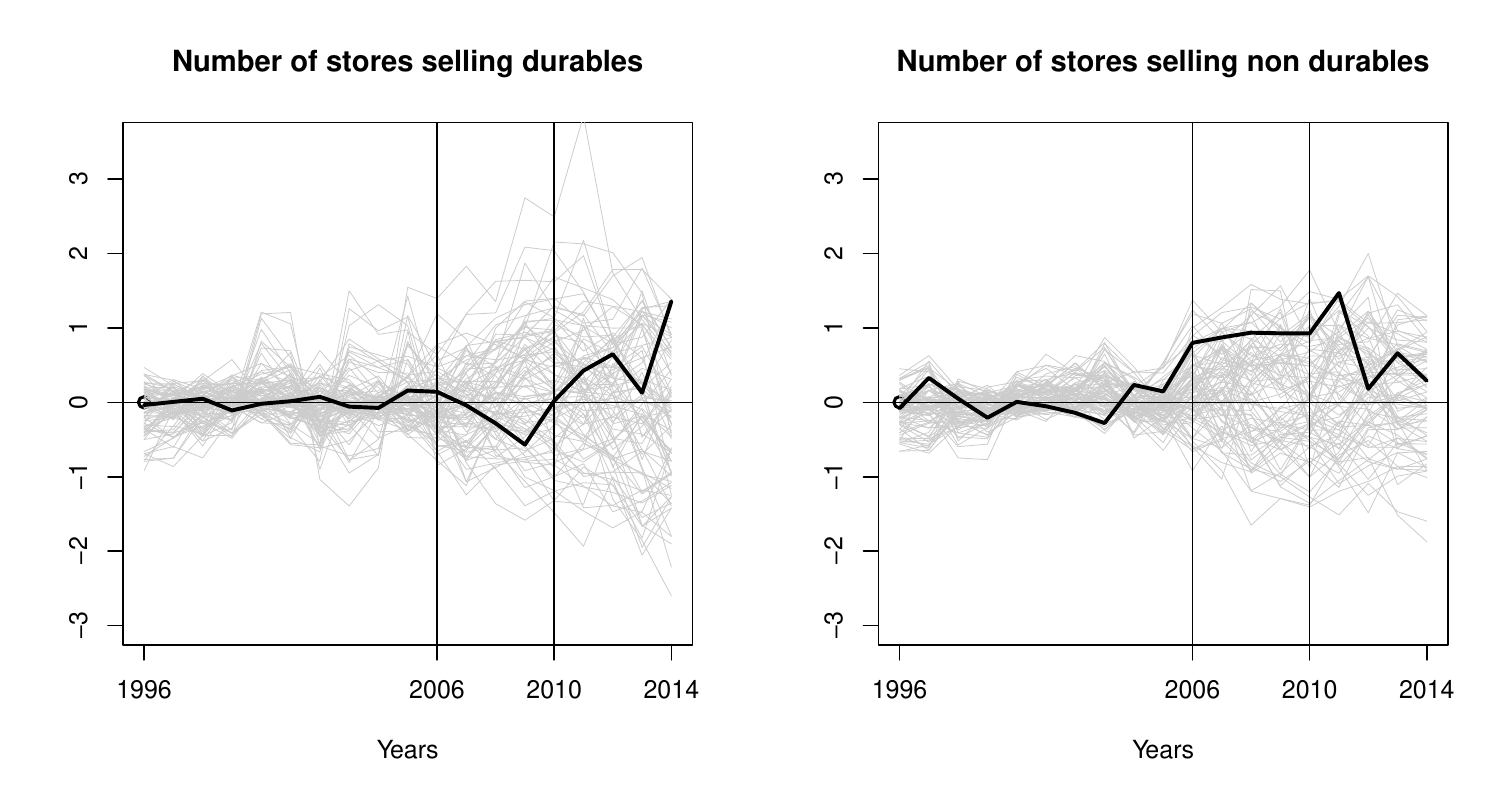}
        \caption{Placebo average spillover effects for stores selling durable and non durable goods - black line: Average spillover for Legnaia streets}
        \label{fig:enter-label}
    \end{figure}

\FloatBarrier
\subsection{Alternative resampling schemes}
Above and beyond the error resampling schemes described in the main text we also consider the following three alternative \textbf{error resampling approaches}. 
Hereafter, we refer to the error sampling scheme we use as ``Scheme I.''

\begin{figure}
    \centering
    \includegraphics[width=\textwidth]{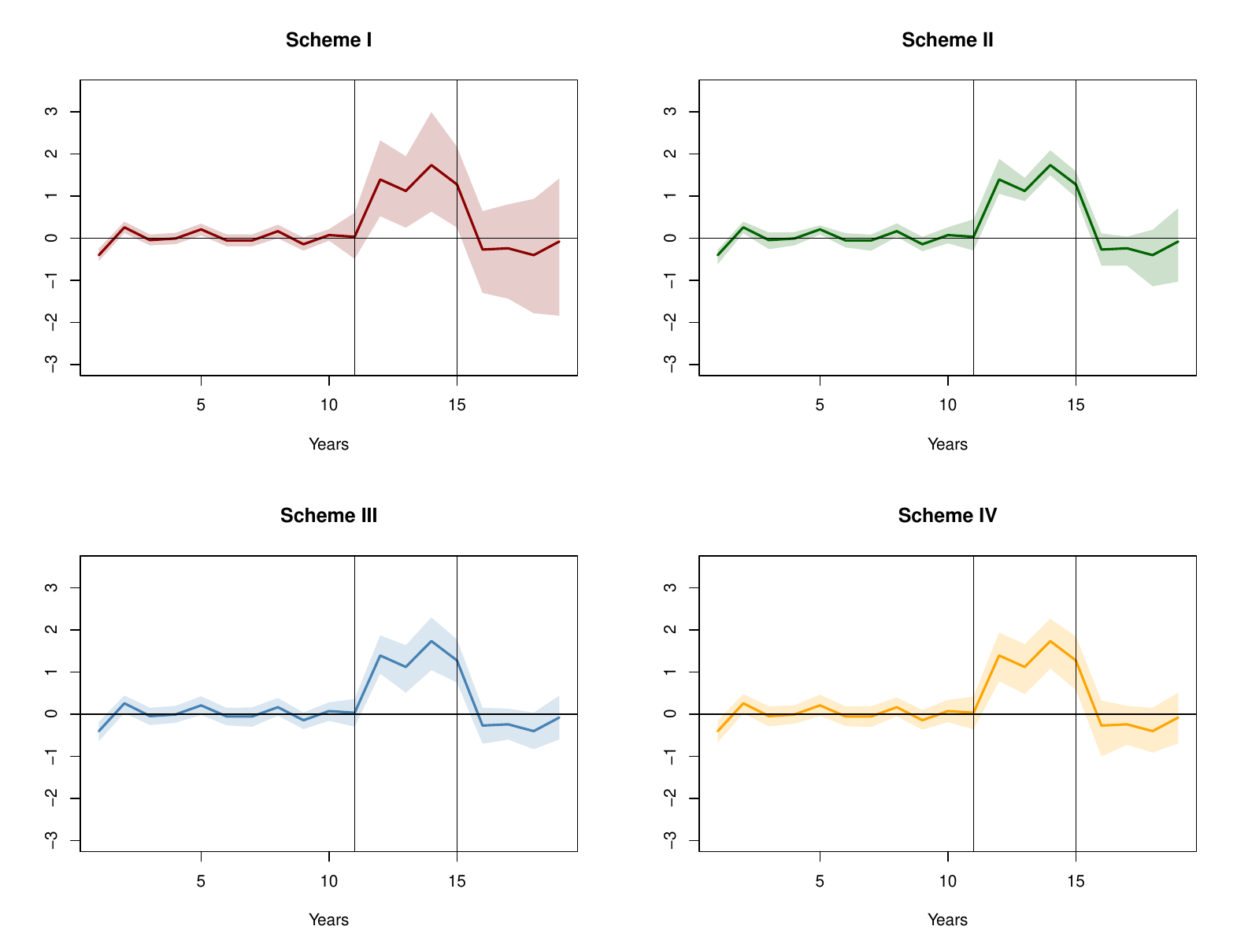}
    \caption{Treatment effect for stores selling durables under the 4 proposed resampling methods, shaded area: 90\% confidence intervals}
    \label{fig:ds_random}
\end{figure}

\begin{figure}
    \centering
    \includegraphics[width=\textwidth]{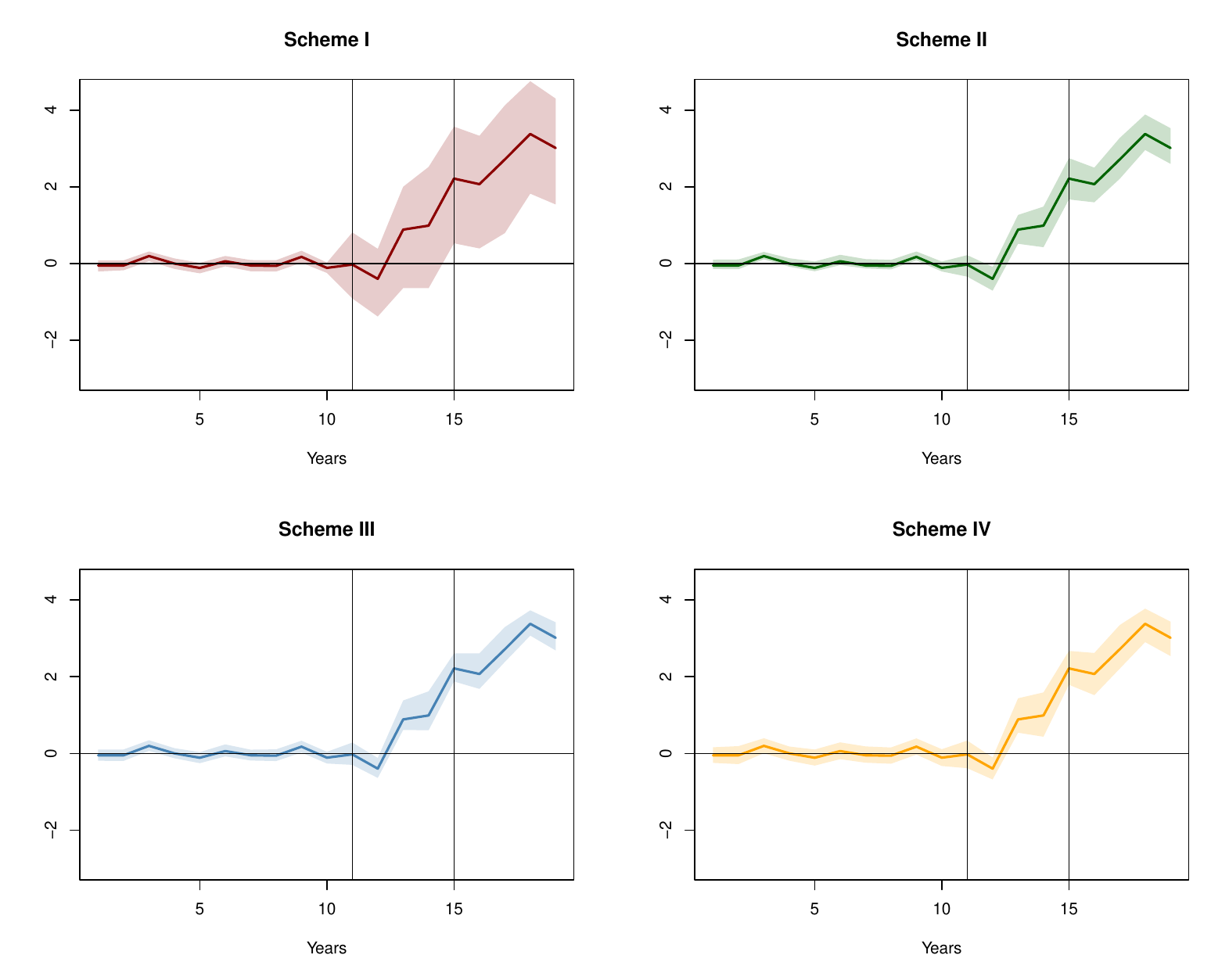}
    \caption{Treatment effect for stores selling non durables under the 4 proposed resampling methods, shaded area: 90\% confidence intervals}
    \label{fig:nds_random}
\end{figure}

\begin{figure}
    \centering
    \includegraphics[width=\textwidth]{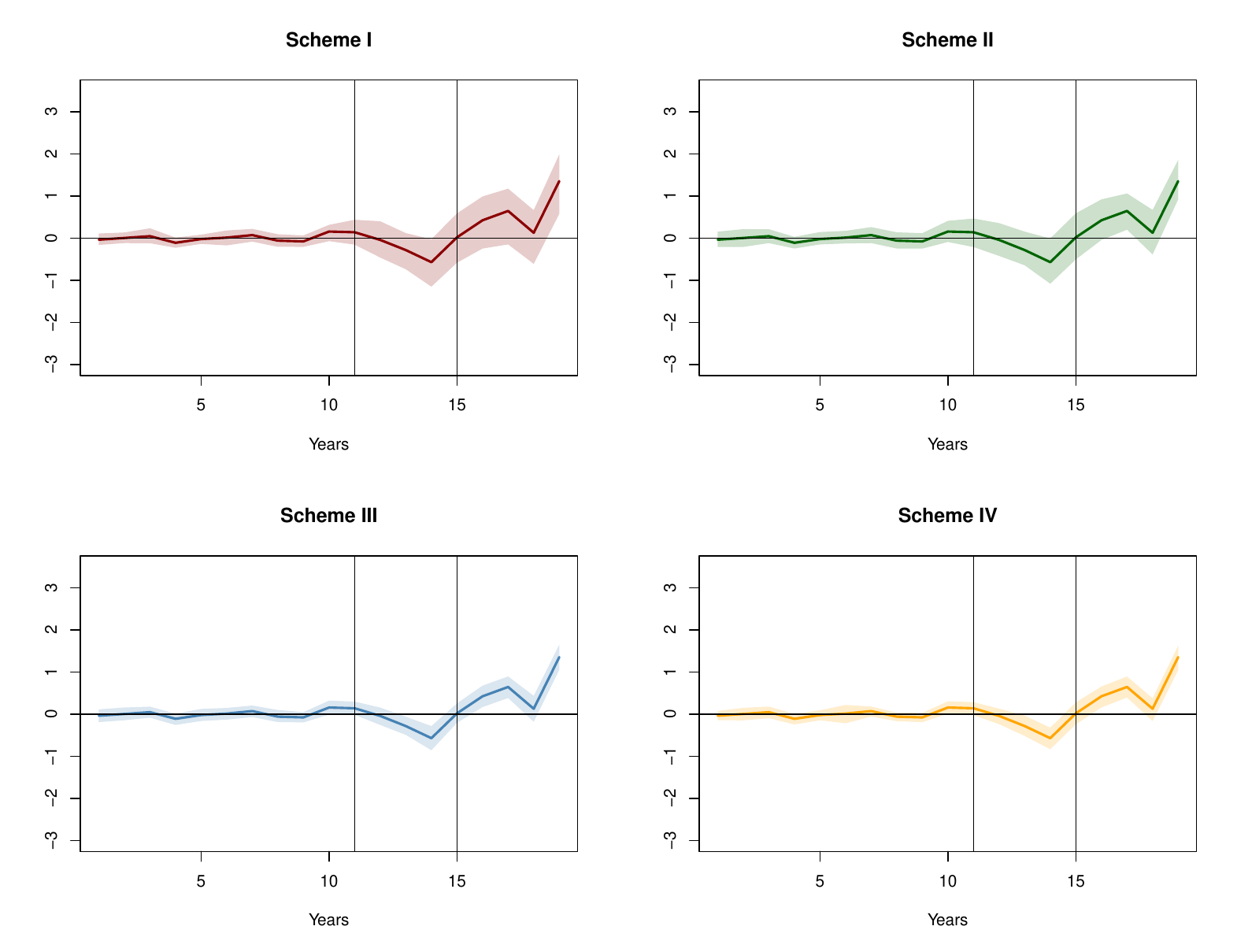}
    \caption{Average treatment effect for stores selling durables under the 4 proposed resampling methods, shaded area: 90\% confidence intervals}
    \label{fig:avg_ds_random}
\end{figure}

\begin{figure}
    \centering
    \includegraphics[width=\textwidth]{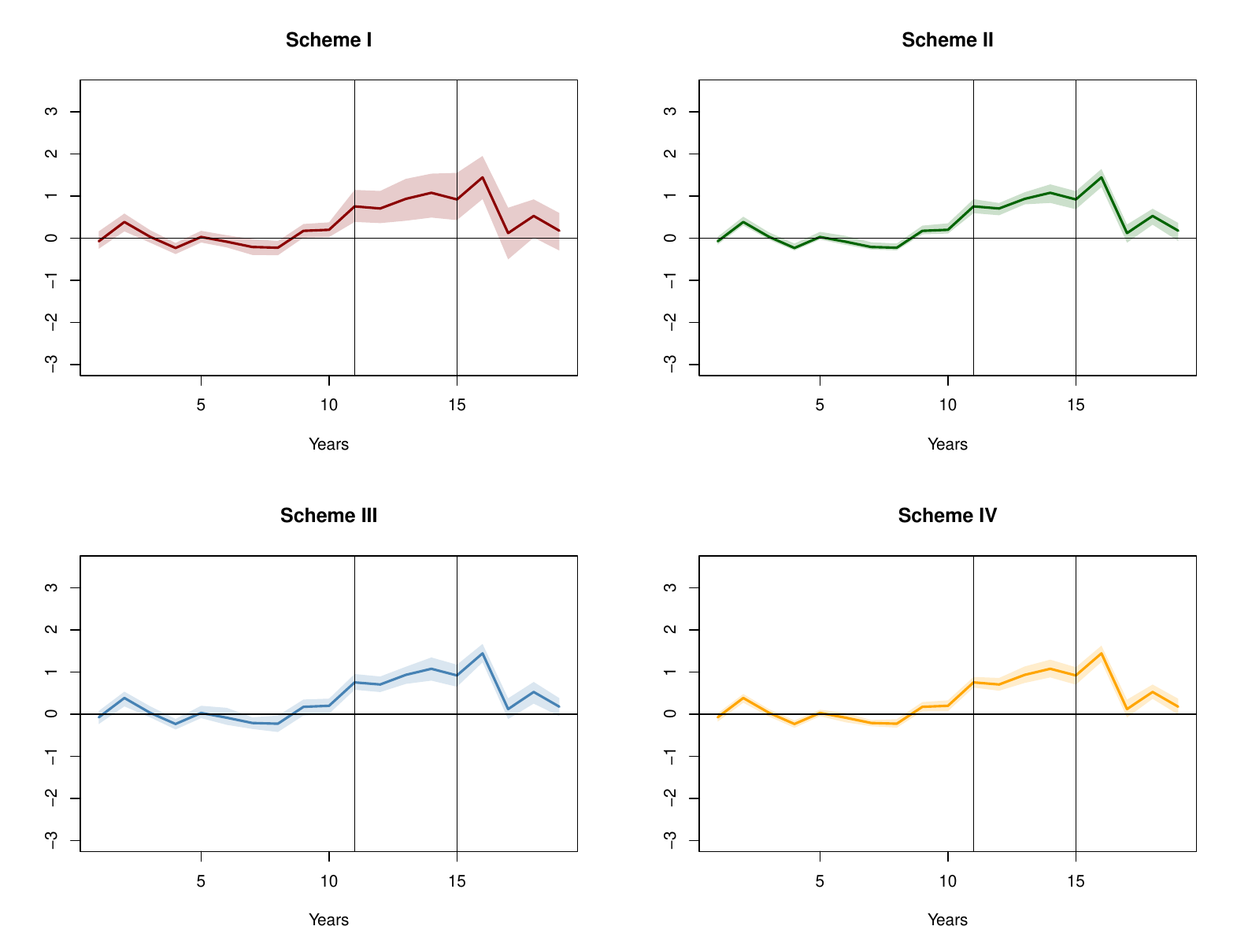}
    \caption{Average treatment effect for stores selling durables under the 4 proposed resampling methods, shaded area: 90\% confidence intervals}
    \label{fig:avg_nds_random}
\end{figure}

\begin{itemize}
    \item Scheme II - In-time error sampling for units in the treated cluster. 
    $(a)$ First, for each $i \in \N_1$ we estimate the vector of pre-treatment errors $\be_i=\left[\be_{i,1}, \dots, \be_{i,T_0} \right]'$ as $\widehat{e}_{i,t}=Y_{i,t} - \widehat{Y}_{i,t}$, with $\widehat{Y}_{i,t}$ estimated using the syntetic control method. 
    
    $(b)$ Then, we resample with replacement a $T_0-$dimensional vector of errors $\be^*_i$ from $\be_i$ and construct pseudo-pre-treament outcomes $Y_{i,t}^* = \widehat{Y}_{i,t} + \be^*_{i,t}$  for  $t=1, \dots, T_0$.
    $(c)$ Finally we re-estimate direct and spillover effects
     using  pseudo-pre-treament outcomes $Y_{i,t}^*$, $i \in \N_1$, $t=1, \ldots, T_0$.
    $(d)$ We repeat step $(b)$ and $(c)$ $B$ times  (in our case 1000 times), and construct pointwise biased corrected confidence intervals from the resampling distribution of the pseudo-effects $\tau_{1,t}^*$, and $\widehat{\delta}^{\N_1,*}_t, $ following the pivotal method.
    \item Scheme III - In-time error sampling for all units.
    
    
    We implement a resampling scheme similar to the in-time error sampling scheme for units in the treated cluster, but in addition to resampling the pre-treatment errors for the units belonging to the treated cluster, we also resample the pre-treatment errors for the control units and constructing pseudo-pre-treatment outcomes for all units.
    It is worth noting that for each unit we resample pre-treatment errors from its own error series, as in Scheme I. 
   
    \item Scheme IV -  In-time error sampling scheme for units in the treated cluster and mixed error sampling scheme for control units.

    We use the in-time error sampling scheme for units in the treated cluster (Scheme II). For control units, we construct pseudo-outcomes $Y^\ast_{i,t}$, $i=N_1+1, \ldots, 1+N$, $t=1, \ldots, T$ by resampling with replacement the error terms
    $\widehat{e}_{i,t} = Y^\ast_{i,t}-\widehat{Y}_{i,t}$ both over time and across units. Therefore, in this sampling scheme, we lose the temporal order of the error terms also for control units, which instead maintain under sampling scheme I.

\end{itemize}
Figures \ref{fig:ds_random}, \ref{fig:nds_random}, \ref{fig:avg_ds_random} and \ref{fig:avg_nds_random} show the 90\% pointwise biased corrected confidence intervals we obtain using the four error sampling scheme described above

We find that the alternative schemes grant similar results. It is worth noting that among the proposed inferential procedure, the sampling scheme I, the approach we use, is the most conservative leading to the widest confidence intervals, even if it provides similar evidence in terms of statistical significance to the other error sampling schemes.

\FloatBarrier
\section{Tables}
\begin{table}
      \caption{Values for the outcomes of interest for the streets in the treated neighborhood}
      \label{tab:descrittive}%
\resizebox{0.55\textwidth}{!}{
    \begin{tabular}{|c|c|ccccc|}
    \hline
    \multicolumn{7}{|c|}{\textbf{Number of stores selling durable goods}} \\
    \hline
          & \textbf{Talenti} & Pollaiolo & Pisana & Scandicci & Magnolie & Baccio\\
          \hline
    1996  & \textbf{7.983} & 13.524 & 10.007 & 6.299 & 10.448 & 7.092 \\
    1997  & \textbf{9.166} & 15.026 & 11.042 & 6.693 & 10.448 & 7.092 \\
    1998  & \textbf{8.870} & 13.899 & 10.352 & 5.906 & 10.448 & 7.447 \\
    1999  & \textbf{9.758} & 13.899 & 10.697 & 6.299 & 11.194 & 7.801 \\
    2000  & \textbf{11.236} & 15.402 & 11.042 & 6.693 & 12.687 & 8.511 \\
    2001  & \textbf{11.236} & 15.402 & 11.387 & 7.874 & 12.687 & 9.220 \\
    2002  & \textbf{10.053} & 13.148 & 10.697 & 7.087 & 11.940 & 8.511 \\
    2003  & \textbf{10.053} & 12.772 & 10.697 & 7.874 & 11.194 & 9.929 \\
    2004  & \textbf{9.758} & 12.772 & 10.007 & 6.693 & 11.940 & 10.638 \\
    2005  & \textbf{9.758} & 12.772 & 11.042 & 6.299 & 12.687 & 12.057 \\
    2006  & \textbf{9.462} & 12.772 & 11.042 & 6.693 & 12.687 & 12.057 \\
    2007  & \textbf{10.053} & 13.148 & 11.732 & 5.512 & 11.194 & 12.057 \\
    2008  & \textbf{9.758} & 11.270 & 11.732 & 5.512 & 9.701 & 11.348 \\
    2009  & \textbf{9.758} & 10.894 & 11.732 & 5.118 & 8.955 & 11.348 \\
    2010  & \textbf{9.758} & 12.397 & 11.387 & 5.118 & 10.448 & 11.702 \\
    2011  & \textbf{8.575} & 12.021 & 12.077 & 5.906 & 10.448 & 12.057 \\
    2012  & \textbf{7.688} & 12.397 & 11.732 & 6.299 & 11.194 & 10.638 \\
    2013  & \textbf{7.688} & 11.645 & 9.662 & 6.299 & 11.194 & 9.574 \\
    2014  & \textbf{7.096} & 12.021 & 9.662 & 7.480 & 11.194 & 8.865 \\
    \hline
    \multicolumn{7}{|c|}{\textbf{Number of stores selling non-durable goods}} \\
    \hline
          & \textbf{Talenti} & Pollaiolo & Pisana & Scandicci & Magnolie & Baccio \\
          \hline
    1996  & \textbf{6.801} & 9.767 & 10.697 & 6.299 & 11.194 & 8.156 \\
    1997  & \textbf{7.392} & 10.518 & 12.422 & 7.087 & 11.940 & 9.220 \\
    1998  & \textbf{7.392} & 10.143 & 10.007 & 5.512 & 11.194 & 7.801 \\
    1999  & \textbf{7.688} & 10.894 & 10.697 & 5.906 & 11.940 & 8.156 \\
    2000  & \textbf{8.575} & 11.270 & 11.387 & 6.299 & 11.940 & 9.929 \\
    2001  & \textbf{9.758} & 11.270 & 11.732 & 5.906 & 12.687 & 10.993 \\
    2002  & \textbf{7.392} & 9.016 & 10.697 & 5.118 & 11.940 & 9.220 \\
    2003  & \textbf{7.983} & 8.640 & 10.697 & 5.512 & 11.194 & 9.220 \\
    2004  & \textbf{7.688} & 8.640 & 9.662 & 5.512 & 12.687 & 9.929 \\
    2005  & \textbf{7.392} & 9.391 & 11.042 & 5.906 & 12.687 & 10.993 \\
    2006  & \textbf{7.983} & 10.143 & 12.077 & 5.906 & 14.925 & 10.638 \\
    2007  & \textbf{7.688} & 10.894 & 12.077 & 4.724 & 14.925 & 11.348 \\
    2008  & \textbf{7.688} & 10.518 & 12.077 & 4.724 & 14.925 & 10.638 \\
    2009  & \textbf{7.688} & 11.645 & 12.077 & 4.724 & 14.179 & 11.348 \\
    2010  & \textbf{9.166} & 12.397 & 11.387 & 5.118 & 14.179 & 11.702 \\
    2011  & \textbf{9.462} & 12.021 & 12.767 & 5.512 & 14.179 & 12.057 \\
    2012  & \textbf{9.758} & 10.518 & 10.697 & 5.118 & 12.687 & 11.348 \\
    2013  & \textbf{9.462} & 8.640 & 10.697 & 4.724 & 11.940 & 10.284 \\
    2014  & \textbf{8.279} & 7.137 & 8.972 & 4.331 & 11.194 & 9.929 \\
    \hline
    \end{tabular}%
    }
  \end{table}%

\begin{table}
    	\caption{Weights through which the synthetic control values of the outcome variables $Y_{1,t}(1, \bz_{N_1-1})$ and $Y_{i,t}(\bz_{N_1})$. A: number of stores selling durables; B: number of stores selling non-durables
}
	\label{tab:weights}
  \centering
    	\resizebox{\textwidth}{!}{
    \begin{tabular}{|r|cc|cccccccccc|}
    \hline
    & \multicolumn{2}{c}{\textbf{$Y_{1,t}(1, \bz_{N_1-1})$}} & \multicolumn{10}{|c|}{\textbf{$Y_{i,t}(\bz_{N_1})$}}\\
    \hline
          & \multicolumn{2}{c|}{Talenti} & \multicolumn{2}{c}{Pollaiolo} & \multicolumn{2}{c}{Pisana} & \multicolumn{2}{c}{Scandicci} & \multicolumn{2}{c}{Magnolie} & \multicolumn{2}{c|}{Baccio da M.} \\
          \hline
          & A     & B     & A     & B     & A     & B     & A     & B     & A     & B     & A     & B \\
    Affrico & 0     & 0     & 0     & 0 & 0     & 0     & 0     & 0.2703 & 0.0520 & 0.0483 & 0.2103 & 0 \\
    Alderotti & 0     & 0     & 0     & 0     & 0     & 0     & 0     & 0     & 0.0719 & 0     & 0     & 0 \\
    Aretina & 0     & 0.1335 & 0     & 0.1097 & 0     & 0.0521 & 0     & 0     & 0     & 0.0180 & 0     & 0 \\
    Baracca & 0     & 0     & 0     & 0     & 0     & 0     & 0     & 0     & 0     & 0     & 0     & 0 \\
    Caracciolo & 0.115 & 0     & 0.1292 & 0.0542     & 0.02  & 0.0834 & 0.184 & 0.1140 & 0     & 0     & 0     & 0 \\
    Centostelle & 0.0678 & 0     & 0     & 0     & 0     & 0     & 0     & 0     & 0     & 0     & 0     & 0 \\
    Corsica & 0     & 0     & 0     & 0     & 0     & 0     & 0.2531 & 0     & 0     & 0     & 0     & 0.2702 \\
    DAnnunzio & 0.1306 & 0     & 0     & 0     & 0     & 0     & 0     & 0     & 0     & 0     & 0     & 0 \\
    Datini & 0.0454 & 0.0344 & 0.1587 & 0     & 0     & 0     & 0     & 0     & 0     & 0     & 0     & 0 \\
    DeSantis & 0     & 0     & 0     & 0 & 0     & 0     & 0     & 0     & 0     & 0.0233 & 0     & 0 \\
    Europa & 0     & 0     & 0.1235 & 0     & 0     & 0     & 0     & 0     & 0     & 0     & 0     & 0 \\
    Faentina & 0     & 0.1742 & 0     & 0     & 0.0182 & 0     & 0     & 0     & 0     & 0     & 0     & 0 \\
    Galliano & 0     & 0     & 0     & 0.3826 & 0     & 0     & 0     & 0.5391 & 0     & 0.0653 & 0     & 0 \\
    Giuliani & 0.0291 & 0     & 0     & 0     & 0     & 0     & 0     & 0     & 0.0449 & 0     & 0     & 0 \\
    Guidoni & 0     & 0     & 0     & 0     & 0     & 0.3818 & 0     & 0.0085 & 0     & 0.3946 & 0     & 0.0376 \\
    Maffei & 0     & 0.0649 & 0     & 0     & 0     & 0     & 0     & 0     & 0     & 0.0164 & 0.0319 & 0.2277 \\
    Maragliano & 0.0085 & 0.1879 & 0.0688 & 0     & 0.3725 & 0     & 0.0572 & 0     & 0     & 0     & 0     & 0 \\
    Mariti & 0     & 0     & 0     & 0     & 0     & 0     & 0     & 0     & 0     & 0.1349 & 0.198 & 0.3335 \\
    Masaccio & 0     & 0     & 0     & 0     & 0     & 0     & 0.2861 & 0     & 0     & 0     & 0     & 0.0668 \\
    Mille & 0.1372 & 0     & 0     & 0     & 0     & 0     & 0     & 0     & 0.4193 & 0     & 0     & 0 \\
    Morgagni & 0     & 0.2263 & 0.1424 & 0     & 0.0177 & 0     & 0     & 0.0681 & 0     & 0.1089 & 0     & 0 \\
    Novoli & 0     & 0     & 0     & 0     & 0     & 0     & 0     & 0     & 0     & 0     & 0     & 0 \\
    Panche & 0     & 0     & 0     & 0.0109 & 0.0707 & 0     & 0     & 0     & 0     & 0     & 0     & 0 \\
    Peretola & 0     & 0     & 0     & 0     & 0     & 0     & 0     & 0     & 0     & 0     & 0     & 0 \\
    Piagentina & 0     & 0     & 0     & 0     & 0     & 0     & 0     & 0     & 03 & 0     & 0     & 0 \\
    Pistoiese & 0     & 0     & 0     & 0     & 0     & 0     & 0     & 0     & 0     & 0     & 0     & 0 \\
    PontealleMosse & 0     & 0     & 0     & 0     & 0     & 0     & 0     & 0     & 0.1295 & 0     & 0.1942 & 0 \\
    PontediMezzo & 0     & 0     & 0     & 0     & 0     & 0.3168 & 0     & 0     & 0     & 0     & 0     & 0 \\
    Pratese & 0.2016 & 0     & 0     & 0     & 0     & 0     & 0     & 0     & 0.1521 & 0     & 0     & 0 \\
    Redi  & 0     & 0     & 0     & 0     & 0     & 0     & 0     & 0     & 0.1132 & 0     & 0.3657 & 0.1018 \\
    Ripoli & 0     & 0     & 0     & 0     & 0     & 0     & 0.1636 & 0     & 0     & 0     & 0     & 0 \\
    Romito & 0     & 0     & 0.1554 & 0.3063 & 0     & 0     & 0     & 0     & 0     & 0     & 0     & 0 \\
    Rondinella & 0     & 0     & 0     & 0     & 0     & 0.1389 & 0     & 0     & 0.0072 & 0     & 0     & 0.0217 \\
    Tavanti & 0     & 0     & 0.0353 & 0.0407 & 0     & 0     & 0     & 0     & 0     & 0     & 0     & 0 \\
    Toselli & 0.2648 & 0.0407 & 0     & 0     & 0     & 0     & 0     & 0     & 0     & 0     & 0     & 0 \\
    Villamagna & 0     & 0.0357 & 0     & 0.0798 & 0.5009 & 0     & 0     & 0     & 0     & 0.1867 & 0     & 0 \\
    VittorioEmanuele & 0     & 0.1289 & 0.1867 & 0 & 0     & 0     & 0.056 & 0     & 0     & 0     & 0     & 0 \\
    Volta & 0     & 0     & 0     & 0     & 0     & 0     & 0     & 0     & 0     & 0     & 0     & 0 \\
    \hline
    \end{tabular}%
    }

\end{table}%

\begin{table}

\caption{Street length}
\centering
\begin{tabular}[t]{lr}
\toprule
   \textbf{Street name} & \textbf{Length (in metres)}\\
\midrule
 Talenti & 1691\\
\hline
Pollaiolo & 1331\\

 Pisana & 1449\\

 Scandicci & 1270\\

 Magnolie & 670\\

 BacciodaMontelupo & 1410\\
\hline
Alderotti & 866\\

Giuliani & 1690\\

 Morgagni & 1061\\

 Panche & 1131\\

 Corsica & 741\\

 Mariti & 618\\

PontediMezzo & 492\\

 Romito & 1548\\

 Tavanti & 1576\\

 VittorioEmanuele & 1708\\

 Baracca & 1958\\

Guidoni & 2375\\

 Novoli & 1782\\

Europa & 2221\\

 Datini & 1336\\

Ripoli & 1751\\

Villamagna & 1478\\

Centostelle & 1182\\

Mille & 1037\\

 Volta & 962\\

Masaccio & 1272\\

 Affrico & 1344\\

 DAnnunzio & 1171\\

 Rondinella & 1151\\

 Aretina & 1607\\

 DeSantis & 539\\

 Piagentina & 509\\

 Galliano & 1237\\

Maragliano & 1083\\

 PontealleMosse & 1366\\

 Redi & 1235\\

 Toselli & 1457\\

Peretola & 2298\\

 Pistoiese & 1430\\

 Pratese & 1417\\

Caracciolo & 758\\

 Faentina & 1770\\

Maffei & 721\\
\bottomrule
\end{tabular}
\end{table}

\FloatBarrier

\end{document}